\documentclass[journal ]{new-aiaa}
\usepackage[utf8]{inputenc}
\usepackage{textcomp}
\usepackage{graphicx}
\usepackage{amsmath}           
\usepackage{amssymb}           
\usepackage{ulem}
\usepackage[version=4]{mhchem}
\usepackage{siunitx}
\usepackage{longtable,tabularx}
\usepackage{caption}
\usepackage{float}
\usepackage{changepage}  
\usepackage{subcaption}

\title{Load Limiting Control for Component Life Extension}

\author{Chams Eddine Mballo\footnote{PhD Candidate, Daniel Guggenheim School of Aerospace
Engineering (Corresponding Author)}, Robert Walters\footnote{PhD Candidate, Daniel Guggenheim School of Aerospace
Engineering}, and Jonnalagadda V.R. Prasad  \footnote{Professor, Daniel Guggenheim School of Aerospace
Engineering, AIAA Fellow.}}
\affil{Georgia Institute of Technology, Atlanta, Georgia, 30332}

\begin{document}
\maketitle
\begin{adjustwidth}{0.75in}{0.75in}  
\noindent\footnotesize
\textbf{Author Accepted Manuscript (AAM).}\;
Version of Record in \emph{J.\ Guid.\ Control Dyn.},
Vol.\ 48, No.\ 2 (Feb 2025), pp.\ 255–268.
DOI \url{https://doi.org/10.2514/1.G007854}. © 2024 AIAA.
\end{adjustwidth}

\vspace{1.0\baselineskip}  

\begin{abstract}

This paper presents the development of a novel life-extending control scheme for critical helicopter components subjected to significant fatigue loading. The primary objective is to synthesize a more efficient and less conservative life-extending control scheme than those currently available in the literature. The proposed Load Limiting Control (LLC) scheme is a viable solution that addresses several issues that current life-extending control schemes suffer from, such as the neglect of fatigue damage induced by the harmonic component of loads and the inability to distinguish between aggressive and non-aggressive maneuvers. The proposed LLC scheme treats desired harmonic load limits as limit boundaries and recasts the problem of load limiting as a vehicle limit by computing a Control Margin (CM) using a limit detection and avoidance module. The computed CM is used as a cue to the pilot. The limit detection and avoidance module comprises an optimization algorithm, a model predictive controller, and a computationally simple on-board dynamical model. Simulations were conducted to demonstrate the effectiveness of the LLC scheme in limiting harmonic pitch link loads during flight. One significant outcome is that, with sufficient training, the pilot can skillfully track the cue within 0.5 seconds of initiating the tracking task. 

\end{abstract}

\section*{Nomenclature}


{\renewcommand\arraystretch{1.0}
\noindent\begin{longtable*}{@{}l @{\quad=\quad} l@{}}
$A$  & Linear time invariant model state matrix \\
$\hat{A}$ & {Reduced order linear time invariant model state matrix} \\
$B$ &  Linear time invariant model input matrix \\
$\hat{B}$ & {Reduced order linear time invariant model input matrix} \\
$C$& Linear time invariant model output matrix \\
$\hat{C}$&  {Reduced order linear time invariant model output matrix} \\
$D$ & Linear time invariant model direct transition matrix \\
$\hat{D}$&  {Reduced order linear time invariant model direct transition matrix} \\
$F({\psi})$ & Linear time periodic model state matrix \\
$G({\psi})$ & Linear time periodic model input matrix \\
$P({\psi})$ & Linear time periodic model output matrix\\
$p$ &  Body roll rate, deg/s\\
$q$ &  Body pitch rate, deg/s\\
$R({\psi})$ & Linear time periodic model direct transition matrix \\
$r$ &  Body yaw rate, deg/s\\
$T_{p}$ &  Time horizon, s\\
$U$ & Augmented control vector\\
$u$ & Control vector\\
$V$ & Airspeed, kt\\
$v_{x}, v_{y}, v_{z}$ & Velocities in body frame, kt\\
$X$ & Augmented sate vector\\
$x$ & State vector\\ 
$x_{R}$ & State vector of reduced order model\\
$x_{B}$ & Rigid body state vector\\ 
$Y$ & Augmented output vector\\
$y$ & Output vector\\
$y_{harm}$ & Harmonic load, lbs\\
$y_{R}$ & Output vector of reduced order model\\
$\beta_{0}$ & Coning, deg\\
$\beta_{1c}$ & Longitudinal flapping, deg\\
$\beta_{1s}$ & Lateral flapping, deg\\
$\theta$ & Pitch attitude, deg\\
$\phi$ & Roll attitude, deg\\
$\psi$ & Nondimensional time or main rotor azimuth angle, deg\\
$\Delta$ & Perturbation from equilibrium\\
$\delta_{col}$ & Collective input, \%\\
$\delta_{lat}$ & Lateral cyclic input, \%\\
$\delta_{lon}$ & Longitudinal cyclic input, \%\\
$\delta_{ped}$ & pedal input, \%\\

\multicolumn{2}{@{}l}{Subscripts}\\
$()_{0}$ & Average or $0^{th}$ harmonic term\\
$()_{ext}$ & Extremal value\\
$()_{max}$ & Maximum value\\
$()_{nc}$ & $n^{th}$ cosine harmonic term\\
$()_{ns}$ & $n^{th}$ sine harmonic term\\
$()_{n/rev}$ & $n/rev$ magnitude component\\
$()_{trim}$ & Trim value\\

\end{longtable*}}

\section{Introduction}

\lettrine{A} 2012 survey of the past 30 years, conducted within Augusta Westland Limited (AWL) Materials Technology Laboratory, concluded that fatigue failures account for approximately $55\%$ of all premature failures in helicopter components~\cite{Davies}. The causes of low cycle fatigue primarily stem from aircraft maneuvers. Critical helicopter components, classified as Grade-A Vital components by regulatory authorities, are subject to significant fatigue loading, where failure could lead to catastrophic events. A list of fatigue-critical components~\cite{Kaye} on the AH-64A Apache reveals that many Grade-A Vital components are located in the rotor system, posing challenges not only for real-time load monitoring but also for the development of effective control strategies to extend the life of these components.

Current methods for structural health and usage monitoring and Load Alleviation Control (LAC) rely on distributed sensing and operational monitoring to infer usage and estimate fatigue in critical components. Such methodologies involve significant challenges due to the difficulty in placing sensors on rotating components and other hot-spot locations often characterized by maximum stresses. For example, previous work~\cite{Jeram} aimed at limiting pitch link loads has utilized proxy models of the vibratory loading. A classic example is the Equivalent Retreating Indicated Tip Speed (ERITS) parameter. ERITS has been correlated as a function of airspeed and normal load factor with vibratory pitch link loads from retreating blade stall onset and can be used to indirectly constrain the pitch link loads. Additionally, curve fits of pitch link vibratory loads as a function of airspeed have been employed to limit the peak-to-peak pitch link load. Furthermore, various non-physics-based models, developed using statistical methods or neural networks~\cite{add6}, have been used in the synthesis of Health and Usage Monitoring Systems (HUMS). These systems could potentially be used in the development of future control strategies for component life extension~\cite{ref39, ref40, ref41}.  

The use of non-physics-based and proxy models for real-time load monitoring has been a major drawback in the development of effective control strategies for critical component life extension. These models are data-driven and hence require significant amounts of training data, and they also fail to provide the higher-harmonic dynamics of the vibratory loads, a factor essential for fatigue analysis. Therefore, there is a need for high-fidelity physics-based models that can be rapidly constructed offline and can accurately capture, in real-time, the higher harmonic dynamics of the vibratory loads on critical helicopter components during flight. Such models can significantly enhance the design of control schemes for helicopter component life extension.

In an effort to address this gap in the literature, recent studies~\cite{ref9, ref10, ref_olcer} have explored methods for approximating coupled body-rotor-inflow dynamics into a Linear Time Invariant (LTI) form that is suitable for integrated flight/rotor controller development. Such an LTI model captures the higher harmonic dynamics of the vibratory loads. The developed methods use harmonic decomposition to represent higher frequency harmonics as states in an LTI state-space model, and they offer the potential for real-time estimation of the effect of control inputs on component dynamic loads. Such real-time estimation of component level dynamic loads provides the opportunity for real-time monitoring of component loads, and more importantly, the development of control schemes designed to alleviate or limit component dynamic loads by automatically limiting (or providing cues to the pilot regarding) excessively aggressive maneuvers.

The first application of the LTI model of coupled body-rotor-inflow dynamics of~\cite{ref9, ref10, ref_olcer} in the development of life-extending control schemes for critical helicopter components is presented in~\cite{ref_umbi1} and~\cite{ref_umbi2}. In those studies, LAC strategies specifically designed for the UH-60 helicopter are proposed to extend the life of rotating blade root pitch links. The proposed LAC strategies aim to reduce component dynamic loads (e.g., peak-to-peak) and consequently decrease peak-to-peak stresses, potentially leading to a reduction in fatigue life usage. The pursued LAC schemes consider a Linear Quadratic Control (LQR) solution for arriving at feedback control laws that trade between maneuver command following and load alleviation. Consequently, load alleviation is implicitly used in arriving at a compromise set of gains as a trade-off between maneuver performance and load alleviation. The proposed LAC schemes are seen to be effective in reducing the peak-to-peak total pitch link load during maneuvering flight. 

While LAC offers a computationally simple scheme, it leads to a conservative approach for two reasons. First, it is incapable of discerning aggressive from non-aggressive maneuvers. Second, it completely neglects the effect of specific harmonic loads on localized damage (i.e., no distinction is made between different harmonic load effects on accumulated component fatigue). With an LAC design, the trade-off between maneuver performance and component load alleviation is always present, irrespective of the aggressiveness of the maneuver. Furthermore, since LAC is achieved via modifications to the Flight Control System (FCS), it becomes very challenging during flight to track and limit the most damage threatening harmonic loads.

This paper proposes a novel control scheme for critical helicopter component life extension using the LTI model of coupled body-rotor-inflow dynamics. The proposed Load Limiting Control (LLC) for component life extension is a more efficient and less conservative scheme than the traditional LAC scheme since it is able to discern between aggressive and non-aggressive maneuvers and takes into account the impact of harmonic loads on localized damage. The LLC scheme is designed to trade maneuver performance for component load limiting only at the pilot's discretion during aggressive maneuvering. Hence, unlike the LAC scheme, the proposed LLC scheme does make a distinction between aggressive versus non-aggressive maneuvers. The LLC scheme restricts desired harmonic loads to user-selected thresholds, treating these thresholds as limit boundaries to directly reduce the fatigue life usage associated with the harmonic loads. Since the LLC scheme does not rely on modifications to the AFCS, adjusting both targeted harmonic loads and user-selected thresholds during flight becomes easily achievable. This implies that the LLC scheme can be used to track and limit during flight the harmonic load that relates most to damage growth, and consequently, to the life of a component.

The main focus of this work can be summarized as follows. The formulation of a harmonic component load limiting scheme via automatic feedback of estimated control margins for limiting control inputs. Simulation evaluations of the developed scheme for harmonic load limiting via automated limiting of control inputs. The formulation of a load limiting scheme via piloted cueing of the available control margins. Piloted simulation evaluations of the scheme for harmonic component load limiting via pilot cueing.

The paper is organized as follows. First, a brief description of the methodology used for extracting a higher-order LTI model of coupled body-rotor-inflow dynamics is presented. Next, the dimension of the higher-order LTI model is reduced using a model order reduction technique based on the singular perturbation method. Following that, the mathematical formulation of the proposed LLC scheme using the reduced-order LTI model is presented. Finally, the effectiveness of the LLC scheme is assessed in batch and real-time piloted simulations. The paper concludes with remarks summarizing the main points of the study. 


\section{Analytical Formulation of Higher Order LTI Model}

The extraction of a higher order LTI model from a high-fidelity nonlinear helicopter model is presented in this section. 

\subsection{Nonlinear model}

This research utilizes FLIGHTLAB$\textsuperscript{\textregistered}$~\cite{ref13} to develop a high-fidelity nonlinear model of a generic single main rotor helicopter, comparable in size and weight to the UH-60. The nonlinear helicopter model weighs approximately 17,000 lbs, has a variable pitch horizontal tail, and features an articulated rotor system with elastic blades. To accurately represent rotor loads, the helicopter model integrates flexible blades encompassing in-plane, out-of-plane, and torsional bending modes. Additionally, it incorporates a 33-state Peters-He dynamic inflow model, comprehensive non-linear aerodynamic look-up tables for airframe and rotor blade aerodynamic coefficients, along with models for the swashplate actuator, tail rotor actuator, and the Bailey tail rotor.

For the simulations considered in this study, the horizontal tail angle is kept fixed at $0^\circ$. The horizontal tail is not used to trim the aircraft, nor is any scheduling with flight conditions performed to adjust its position. While the horizontal tail does affect the trim load, the primary focus of this work is on the use of the proposed LLC scheme to limit transient loads rather than trim loads.

\subsection{Higher-order LTI model}

Following the method described in~\cite{ref_lopez}, an LTI model using harmonic decomposition of Linear Time Periodic (LTP) states, with a first order representation (i.e., separate displacement and velocity states), is developed from the aforementioned nonlinear model. 

Considering an LTP model as defined in Eqs.~(1) and (2), the process of harmonic decomposition for extracting an LTI model relies on the approximation of the state vector, $x$, as outlined in Eq.~(3). 

 \begin{equation} \dot{x}=F(\psi)x +G(\psi)u \label{eq:1} \end{equation} 

 \begin{equation} y=P(\psi)x +R(\psi)u \label{eq:2} \end{equation}

\begin{equation} x \approx x_{0}+ \sum_{n=1}^{N} [x_{nc}cos(n\psi) +x_{ns}sin(n\psi) ]
\label{eq:2} \end{equation} 

\noindent In the equations above, $x_{0}$ represents the average component of $x$, while $x_{nc}$ and $x_{ns}$ correspond, respectively, to the n/rev cosine and sine harmonic components of $x$. $\psi$ is a nondimensional time parameter known as the azimuth angle, related to time through the rotor's angular velocity ($\Omega$).

 \begin{equation} \psi=\Omega t \label{eq:5} \end{equation}

\noindent A similar approximation is applied to the control ($u$) and the output ($y$) vectors.

\begin{equation} u \approx u_{0}+ \sum_{m=1}^{M} [u_{mc}cos(m\psi) +u_{ms}sin(m\psi) ] \label{eq:2}
\end{equation} 

\begin{equation} y \approx y_{0}+ \sum_{l=1}^{L} [y_{lc}cos(l\psi) +y_{ls}sin(l\psi) ] \label{eq:2} \end{equation}

\noindent $u_{0}$ represents the average component of $u$, while $u_{mc}$ and $u_{ms}$ correspond to the $m^{th}$ harmonic cosine and sine components of $u$. The equation describing the approximation of the output vector, $y$, follows a similar nomenclature\textcolor{red}{.}

An LTI approximation of the LTP model can be obtained by substituting harmonic expansions~\cite{ref_lopez} for $x$, $u$ and $y$, that is, Eqs. (3), (5), and (6) into Eqs. (1) and (2). The resulting equations can then be represented in state-space matrix form by defining augmented state, control, and output vectors as  \begin{equation} X= [x_{0}^T..x_{ic}^T \; \; x_{is}^T..x_{jc}^T \; \;x_{js}^T.. ]^T
\label{eq:2} \end{equation} 

 \begin{equation} U= [u_{0}^T..u_{mc}^T \; \; u_{ms}^T.. ]^T
\label{eq:2} \end{equation} 

 \begin{equation} Y= [y_{0}^T..y_{lc}^T \; \; y_{ls}^T.. ]^T
\label{eq:2} \end{equation} 

\noindent The state-space representation of the resulting LTI model is 

\begin{equation} \dot{X}= [A]X+[B]U
\label{eq:2} \end{equation} 

\begin{equation} Y= [C]X+[D]U
\label{eq:2} \end{equation} 

Detailed expressions for the LTI model matrices $A$, $B$, $C$ and $D$ are developed in~\cite{ref9} and~\cite{ref_lopez}. The LTP model, extracted through linearization from the nonlinear model at a trim flight of 120 knots, includes 8 body states, 33 inflow states (Peters-He Finite state inflow with 4 harmonics and a maximum radial variation power of 8), and 48 multi-blade coordinate (MBC) rotor states that include rigid flap, rigid lead-lag and coupled elastic modes. Thus, the total number of LTP states is 89. Each of these LTP states is then decomposed into 0-8/rev harmonic components, resulting in 1513 total LTI model states. It should be noted that not all 0-8 harmonics may be required to achieve acceptable fidelity in the LTI model~\cite{ref9, ref_lopez}. The input vector of the LTP model is composed of 4 control effectors: longitudinal cyclic, lateral cyclic, main rotor collective, and tail rotor collective. Unlike the state vector, only the zeroth harmonic of the input vector is retained.

This study focuses on a high-speed flight regime at 120 knots due to the significant generation of pitch link loads during maneuvers at this airspeed~\cite{ref_curtiss, ref_voskuijil}. In fact, several studies have been dedicated to developing load alleviation and damage mitigation strategies specifically tailored for this airspeed~\cite{ref11, ref17, ref_daniel}.


The proposed LLC scheme requires the use of an onboard dynamical model for real-time computation of control margin. This computation is executed through an optimization problem initialized using the current states of the nonlinear model that are shared with the onboard dynamical model. This implies that the LLC scheme requires real-time access to these states. While obtaining measurements of states associated with vehicle speed, body rates, and attitudes is common, the same cannot be said for higher-order rotor and inflow states. Therefore, it is crucial for the onboard model state vector to contain as few higher-order rotor and inflow states as possible to ensure the viability of the proposed scheme.

The onboard dynamical model, while computationally simple, needs to provide a reasonable approximation of the true vehicle dynamics. The LTI model developed in this section is an ideal candidate for the onboard model. However, its composition of higher-order rotor and inflow states, along with its very large size, which is directly related to the number of harmonic states, inputs, and outputs retained, can pose a problem for real-time implementation of the LLC scheme. Consequently, the reduction of the higher-order LTI model to a reduced-order model using a two-time-scale method, where only the body states and the flapping states are retained, is described next.

\section{Model Order Reduction using Singular Perturbation Method}

In this section, a reduced-order LTI model is obtained using the singular perturbation method. This method is based on timescale separation, where the system dynamics are decomposed into fast and slow modes~\cite{ref14, ref_koko2}. 

This model order reduction technique can be summarized in two steps. First, the dynamic system under consideration is transformed into a singular perturbation system. This step provides a two-time-scale representation of the original system, decomposing it into fast and slow modes. Then, a reduced-order model of the original system is obtained by performing a quasi-steady approximation of the fast dynamics. In this step, it is assumed that the fast states, which have stable dynamics, reach their equilibrium instantaneously with respect to the slow states. Hence, the singular perturbation procedure accurately captures the low-frequency and steady-state parts of the original system while neglecting the high-frequency dynamics~\cite{ref14}.


A two-time-scale representation of the higher order LTI model can be obtained by using the fact that the rotor and inflow dynamics are much faster than the body dynamics. Since the fast and slow modes of the system  are known a priori, the dynamical system is already in a singularly perturbed form. The state vector of the higher-order LTI model is then divided into slow and fast states as 

\begin{align}
    X &= \begin{bmatrix}
           X_{\text{s}} \\
           X_{\text{f}} 
         \end{bmatrix}
         \label{eq:state_decom}
  \end{align}
\noindent where 
  $X_{\text{s}}$~= refers to slow states and $X_{\text{f}}$~=~refers to fast states. 
\noindent Using Eq.~\ref{eq:state_decom}, the following dynamical system is obtained: 
  
\begin{equation}
\begin{bmatrix}
   \dot {X}_{\text{s}}  \\
   \dot {X}_{\text{f}} \\
  \end{bmatrix}=\begin{bmatrix}
   A_{\text{s}} & A_{\text{sf}}  \\
   A_{\text{fs}} & A_{\text{f}} \\
  \end{bmatrix}\begin{bmatrix}
   X_{\text{s}}\\
   X_{\text{f}}\\ 
   \end{bmatrix} + \begin{bmatrix}
   B_{\text{s}}\\
   B_{\text{f}}\\
  \end{bmatrix}U
  \label{eq:2}
\end{equation}

With the higher-order LTI model decomposed into fast and slow modes, a unique reduced-order model can be obtained by assuming that the fast states reach their equilibrium instantaneously with respect to the slow states (i.e., quasi-steady approximation of the fast dynamics). What follows is a derivation of the reduced-order dynamical system and functional relationship that maps the controls and slow states to the limit parameters. The rotating pitch link is selected as the primary component of interest  within the rotating frame. Therefore, specific harmonic components of pitch link loads are selected as the limit parameters. 

Per the assumption that the fast states reach steady-state very quickly,~$\dot {X}_{\mathrm{f}}$ can be set to zero, and an expression for $X_{\mathrm{f}}$ can be obtained.


\begin{equation} A_{\text{fs}}X_{\text{s}}+ A_{\text{f}}X_{\text{f}}+B_{\text{f}}U=0
\label{eq:2} \end{equation} 
\begin{equation} X_{\text{f}}={A}^{-1}_{\text{f}}[-A_{\text{fs}}X_{\text{s}}-B_{\text{f}}U]
\label{eq:2} \end{equation} 

\noindent By substituting the expression for $X_{\text{f}}$ from Eq.~(15) into Eq.~(13), the dynamic equation for the reduced system is obtained.

\begin{equation} \dot{X}_{\text{s}}= [\hat A]X_{\text{s}}+[\hat B]U
\label{eq:2} \end{equation} 
\noindent where 
\begin{equation} \hat A=A_{\text{s}}-A_{\text{sf}}{A}^{-1}_{\text{f}}A_{\text{fs}}
\label{eq:2} \end{equation} 
\begin{equation} \hat B=B_{\text{s}}-A_{\text{sf}}{A}^{-1}_{\text{f}}B_{\text{f}}
\label{eq:2} \end{equation} 

\noindent The output equation can also be expressed in terms of the slow and fast states. 
\begin{equation}
Y=\begin{bmatrix}
   C_{\text{s}} & C_{\text{f}}  \\
  \end{bmatrix}\begin{bmatrix}
   X_{\text{s}}\\
   X_{\text{f}}\\ 
   \end{bmatrix} + \begin{bmatrix}
   D
  \end{bmatrix}U
  \label{eq:2}
\end{equation}

\noindent Substituting the expression for $X_{\text{f}}$ from Eq.~(15) into Eq.~(19), the output equation for the reduced-order system becomes

 \begin{equation} Y= [\hat C]X_{\text{s}}+[\hat D]U
\label{eq:2} \end{equation} 

\noindent where 

\begin{equation} \hat C=C_{\text{s}}-C_{\text{f}}{A}^{-1}_{\text{f}}A_{\text{fs}}
\label{eq:2} \end{equation} 
\begin{equation} \hat D=D-C_{\text{f}}{A}^{-1}_{\text{f}}B_{\text{f}}
\label{eq:2} \end{equation} 

Using the model order reduction  procedure described above, a $10{\text{th}}$-order LTI model is considered. The reduced order model is derived with slow states consisting of $0^{\text{th}}$ harmonic components of body velocities ($v_{x}$, $v_{y}$, $v_{z}$), body angular velocities ($p$, $q$, $r$), body pitch and roll attitudes ($\theta$, $\phi$), and the $0^{\text{th}}$ harmonics of the longitudinal ($\beta_{1c_{0}}$) and lateral flapping ($\beta_{1s_{0}}$). The resultant slow state vector is defined as 

 \begin{equation} X_{s}= [x^T_{B_{o}}\: \: \beta_{1c_{0}}\: \: \beta_{1s_{0}}]^T
\label{eq:2} \end{equation} 

\noindent The obtained $10{\text{th}}$-order model accurately captures the low-frequency cyclic flap mode in addition to the body modes, as part of the slow dynamics.

A study conducted in~\cite{ref15} assessed the fidelity of different reduced-order LTI models for prediction of blade root pitch link loads, vehicle angular rate, and body velocity component responses. The study concluded that the $10{\text{th}}$-order LTI model provided a relatively good representation of the higher-order LTI model in predicting harmonic pitch link loads. Therefore, a $10{\text{th}}$-order LTI model is used in this research for the synthesis of the load limiting controller. 

\section{Load Limiting Control Law} \label{LOAD LIMITING CONTROL LAW}

This section describes the proposed LLC scheme in detail, highlighting its major innovative features. The use of Model Predictive Control (MPC) for estimating control margins associated with vehicle performance limit boundaries was developed in~\cite{ref38}. These control margins are adopted to formulate a load limiting control scheme by making use of a model predictive receding-horizon control approach.

Various works have focused on determining control margins to enable the synthesis of carefree maneuvering systems~\cite{ref_Kashawlic, ref_Kashawlic1, ref_sahani}. In these studies, control margins are computed by determining the sensitivity of the limit parameter with respect to the controls. However, significant issues are associated with using this method to find control margins. One concern is that the method considers the effect of each control input on the limit parameter separately, thereby neglecting the potentially crucial inter-axis coupling effect of the controls on the limit parameter. Another issue is that this method can lead to significant errors when the sensitivity becomes very small. In this paper, the use of an MPC framework instead of the sensitivity-based method is justified by the need for a more robust, comprehensive, and accurate methodology for computing control margins.
      
Given a load limit ($y_{\mathrm{max}}$) that is not to be exceeded, the LLC scheme uses an on-board model representative of the true vehicle dynamics, a cost function defined over a finite time horizon of $T_{\mathrm{p}}$, and an optimizer to compute, at each instant in time, future extremal control input ($u_{\mathrm{ext}}$) that would allow the component harmonic load ($y_{\mathrm{harm}}$) to reach its limit without exceeding it. This process is illustrated in Fig. 1. Thus, the LLC scheme treats the load limit ($y_{\mathrm{max}}$) as a limit boundary and uses MPC to arrive at an optimal control profile that would give rise to a harmonic load response reaching the limit boundary within a time horizon of $T_{\mathrm{p}}$. The shaded area in Fig. 1 gives a graphical representation of the cost function used by the LLC scheme. The calculated extremal control profile is used to determine the quantity known as Control Margin (CM). The control margin is given by the following equation

 \begin{equation} C\!M(\mathrm{k})= u_{ext}(\mathrm{k})-u_{pilot}(\mathrm{k}-1)
\label{eq:CM} \end{equation} 

The control margin is an important quantity that can be conveyed to the pilot in the form of a cue. This informs the pilots, in real-time, about the maximum permissible control deflection before the harmonic load exceeds the maximum limit. The LLC scheme can also function as an automatic load limiting system. In this scenario, the estimates of extremal control input (i.e., allowable control travel estimates) are directly used to automatically constrain the pilot control inputs, ensuring that the selected harmonic load remains within  the desired maximum value. When integrated with a cueing system, the LLC scheme provides the advantage of enabling the pilot to prioritize between load limiting and maneuver aggressiveness. Consequently, the pilot has the option to follow the cue to avoid exceeding the load limit or disregard it in situations where maneuver performance takes precedence, even at the risk of component load limit violations (e.g., in scenarios requiring collision avoidance with obstacles).  


\begin{figure}[H]
\centering
\includegraphics[width=1.0\textwidth]{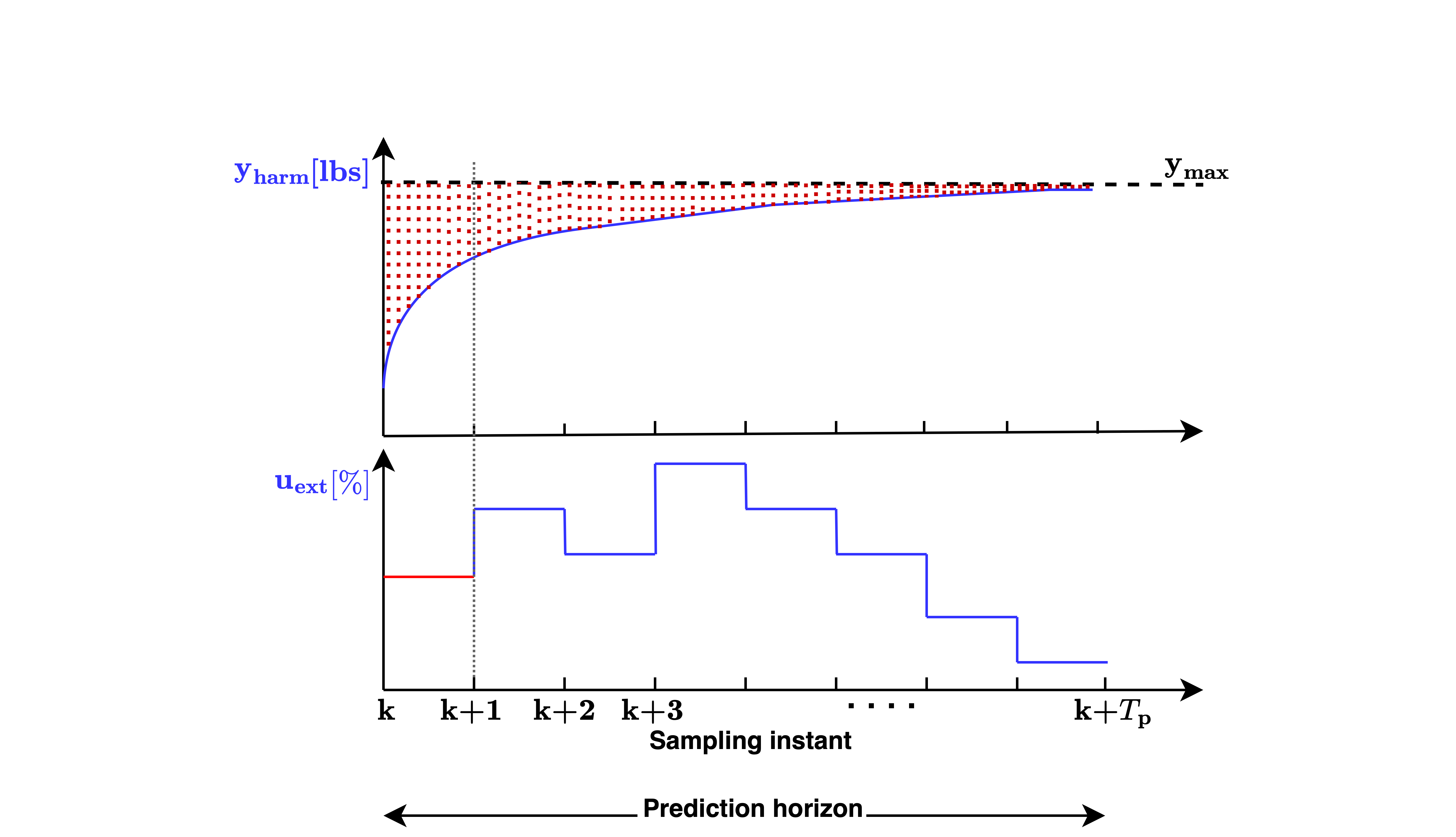}
\caption{LLC scheme: load and extremal control profiles.}
\label{fig:LLC_SCHEME}
\end{figure}

A comparison between the LLC and LAC schemes reveals distinct advantages in the former. While the LAC scheme establishes a more direct trade-off between maneuver performance and load limiting, achieved through the selection of weightings between maneuver command following and resulting harmonic loads, the LLC scheme introduces an indirect trade-off. This indirect trade-off relies on the user-specific value of the harmonic load limit, providing increased flexibility and adaptability. Furthermore, the LLC scheme selectively restrains maneuver performance, limiting it only when at the onset of load limit exceedance. Consequently, non-aggressive maneuvers that pose no risk of load limit exceedance require no trade-off. Unlike the LAC scheme, the LLC scheme offers the capability to limit a component dynamics load within a predefined maximum value ($y_{\mathrm{max}}$). The judicious selection of this maximum value serves as a crucial threshold, preventing significant fatigue damage to the component. An additional key feature of the proposed LLC scheme is its capacity to estimate the future value of the limit parameter. This functionality plays a vital role in the early detection of limit violations, providing pilots with ample time to take preventive actions. The block diagram representation of the LLC scheme via cueing is depicted in Fig.~\ref{fig:LLC_SCHEME}.



\begin{figure}[H]
\centering
\includegraphics[width=1.0\textwidth]{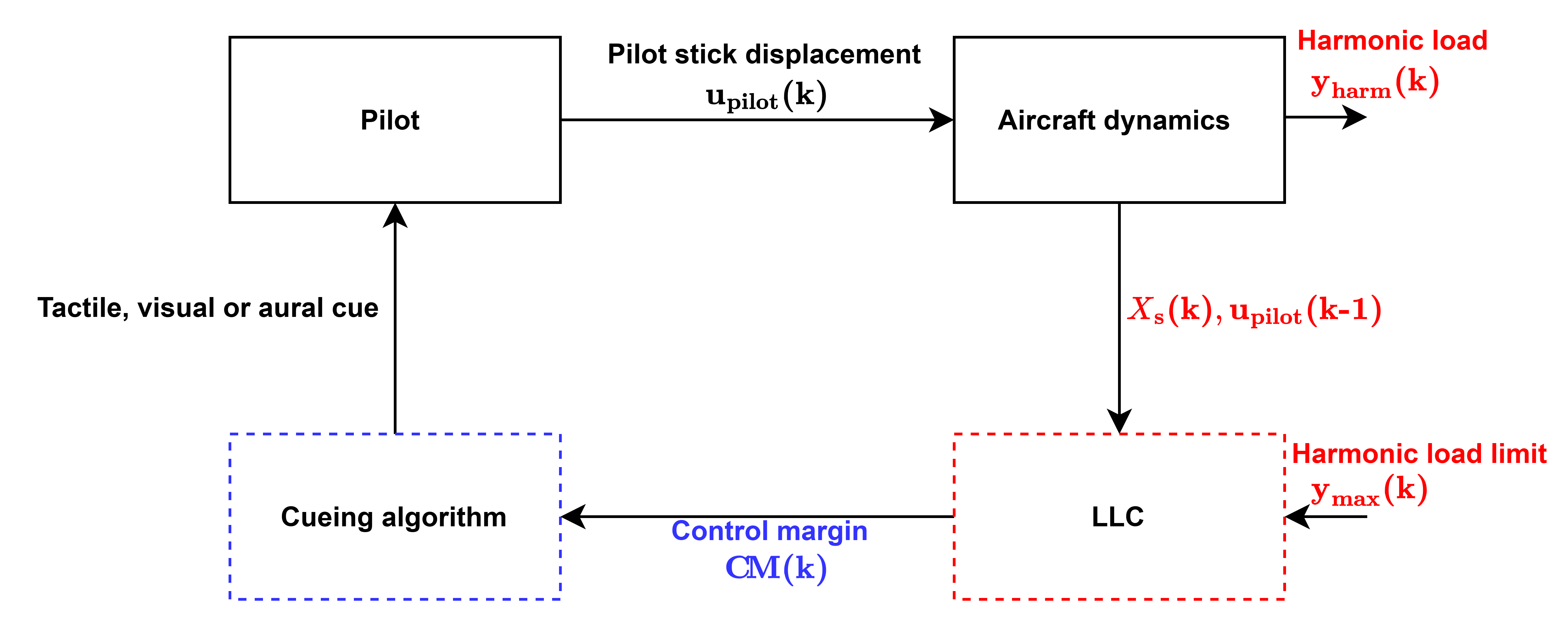}
\caption{LLC scheme via cueing.}
\label{fig:LLC_SCHEME}
\end{figure}


In~Fig.~\ref{fig:LLC_SCHEME}, $X_{\mathrm{s}}(\mathrm{k})$ represents the on-board LTI model state vector at time step $\mathrm{k}$. The LLC scheme assumes that the LTI model states are available from on-board measurements, along with stored values of component harmonic loads at trim. The on-board LTI model uses these measurements to estimate the harmonic loads. $y_{\mathrm{max}}(\mathrm{k})$ is the chosen harmonic load limit at time step $\mathrm{k}$. The pilot has the freedom to change this limit during flight. 

The LLC scheme calculates the extremal control input by solving, at each time instant, a constrained optimization problem with a quadratic cost function, using the $10{\text{th}}$-order LTI model. In particular, the identification of the extremal control associated with the component load limit is formulated as a receding horizon model predictive algorithm. This is expressed in terms of the discrete optimal control problem obtained via the use of the direct transcription method~\cite{BETTS}.





\begin{equation}
\min_{\substack{
U(\mathrm{k}), ...,~ U(\mathrm{k} + T_{\mathrm{p}}-1)\\
X_{\mathrm{s}}(\mathrm{k}+1), ...,~ X_{\mathrm{s}}(\mathrm{k} + T_{\mathrm{p}})
}} [J],~ 
J = \sum_{i=\mathrm{k}}^{\mathrm{k} + T_{\mathrm{p}}} L(\| y_{\mathrm{harm}}(i)\|_{2}, U(i)) \, dt
\label{eq:25_chams}
\end{equation}

subjected to:  \begin{equation} 
\ y_{\mathrm{harm}}(i) \leq y_{\mathrm{max}}(\mathrm{k})
\label{eq:27_chams}, \hspace{0.5cm} i = \mathrm{k}\text{,...,}~\mathrm{k} + T_{\mathrm{p}}  \end{equation} 

\begin{equation} u_{\mathrm{pilot}_{\mathrm{min}}} \leq u_{\mathrm{pilot}}(\mathrm{k}-1) + U(i) \leq u_{\mathrm{pilot}_{\mathrm{max}}}, \hspace{0.5cm} i = \mathrm{k}\text{,...,}~\mathrm{k} + T_{\mathrm{p}}
\label{eq:28_chams} \end{equation}

\begin{equation} X_{\mathrm{s}}(i+1)= [\hat A]_{\mathrm{d}}X_{\mathrm{s}}(i)+[\hat B]_{\mathrm{d}}U(i)
\label{eq:26_chams}, \hspace{0.5cm} i = \mathrm{k}\text{,...,}~\mathrm{k} + T_{\mathrm{p}} - 1 \end{equation} 


\noindent In the optimization problem above, the matrices $[\hat A]_{\mathrm{d}}$ and $[\hat B]_{\mathrm{d}}$ are discrete versions of the $10{\text{th}}$-order state and input matrices, respectively. The cost function integrand is defined as 
 

\begin{equation}
\begin{aligned}
L(\|y_{\mathrm{harm}}(i)\|_{2},U(i)) ={} & \ (\|y_{\mathrm{harm}}(i)\|_{2}-y_{\mathrm{max}}(\mathrm{k}))^TQ(\|y_{\mathrm{harm}}(i)\|_{2}-y_{\mathrm{max}}(\mathrm{k})) + 
       (U(i)-u_{\mathrm{pilot}}(\mathrm{k-1})))^TR(U(i)-u_{\mathrm{pilot}}(\mathrm{k-1}))) \\
\label{eq:29_chams}
\end{aligned}
\end{equation}

\noindent where $R$ and $Q$ are symmetric positive definite matrices of design coefficients. They penalize, respectively, the control activity and the tracking error of the limit parameter in reaching its boundary within the selected time horizon, $T_{p}$. 

The LLC scheme presents a challenge in terms of computational complexity. Solving the optimization problem outlined in Eq.~\ref{eq:25_chams} through Eq.~\ref{eq:29_chams} to determine extremal control inputs or, equivalently, control margin estimates proves to be challenging. This difficulty arises from the potential highly nonlinear nature of the mathematical expression governing the constrained harmonic load ($y_{\mathrm{harm}}$). Leveraging the computational advantages associated with solving convex quadratic optimization problems, a practical strategy for real-time optimization involves initially convexifying the problem, allowing it to be transformed into a convex quadratic program. Subsequently, an ad-hoc solver, such as the CVXGEN solver~\cite{ref_cvx}, can be employed.


As a proof-of-concept, the magnitude of the 1/rev pitch link harmonic load is chosen as the primary harmonic load of interest, given its significant magnitude dominance compared to other harmonic loads. It is crucial to emphasize that, depending on the application, any other harmonic load could be chosen. The expression for the two norm of the 1/rev pitch link harmonic load is given by: 

\begin{equation}  \|y_{1/\mathrm{rev}}(.)\|_{2}= \sqrt{(y_{1c_{(trim)}} +y_{1c}(.))^2 + (y_{1s_{(trim)}} +y_{1s}(.))^2}
\label{eq:30_chams} \end{equation} 

\noindent The limit parameter, $\|y_{1/\mathrm{rev}}(.)\|_{2}$, is the total harmonic load (i.e., trim + change from trim). The optimization problem can be convexified by linearly approximating Eq.~\ref{eq:30_chams} as

\;
\begin{equation}  \|y_{1/\mathrm{rev}}(.)\|_{2} \stackrel{\text{}}{\approx} a + by_{1c}(.) + cy_{1s}(.)
\label{eq:31_chams} \end{equation} 

\noindent where $a$, $b$ and $c$  $\in$ $\mathbb{R}$ are given by

\begin{equation}
a=\sqrt{(y_{1c_{(trim)}})^2 + (y_{1s_{(trim)}})^2} 
\label{eq:30} \end{equation} 
\begin{equation}
b=\frac{y_{1c_{(trim)}}}{\sqrt{(y_{1c_{(trim)}})^2 + (y_{1s_{(trim)}})^2} }
\label{eq:30} \end{equation} 
\begin{equation}
c=\frac{y_{1s_{(trim)}}}{\sqrt{(y_{1c_{(trim)}})^2 + (y_{1s_{(trim)}})^2} }
\label{eq:30} \end{equation} 

The MPC formulation, as expressed by~Eqs.~\ref{eq:25_chams}-~\ref{eq:29_chams}, utilizes~Eq.~\ref{eq:31_chams} as a representation of the magnitude of 1/rev harmonic pitch link load. The solution to the optimal control problem provides estimates of the control margin associated with 1/rev pitch link load limit. 

\section{Results}

In this section, the assessment focuses on the fidelity of the extracted higher-order and $10^{\mathrm{th}}$ LTI models in the frequency domain. Subsequently, an investigation into the performance of the proposed load limiting controller is presented, utilizing nonlinear model simulations at 120 knots forward flight. Specifically, limiting the magnitude of the 1/rev pitch link load is considered. This study encompasses both batch and piloted simulations.~Batch simulations were executed to evaluate the performance of the LLC scheme in an ideal and well-controlled environment. Additionally, pilot-in-the-loop tests were conducted to assess the effectiveness of the proposed scheme in a more realistic environment and to obtain comments from the pilot about their interaction with the LLC scheme through the cues.~For the proof-of-concept study considered here, the prediction horizon ($T_{\mathrm{p}}$) is set at 0.0065 seconds, and the maximum load ($y_{\mathrm{max}}$) is held constant at 350 lbs. It is important to emphasize that the selected value of $y_{\mathrm{max}}$ (i.e., 350 lb.) is arbitrary and does not represent the true maximum load threshold. As mentioned earlier, $y_{\mathrm{max}}$ is user-selected and can be easily changed during flight. Various other simulations scenarios involving different harmonic loads and load limits were also considered. For the sake of brevity, this paper primarily focuses on presenting one specific set of results where the 1/rev pitch link load is limited to 350 lbs.

\subsection{LTI model validation}

\begin{figure}
     \centering
     \begin{subfigure}[b]{0.56\textwidth}
         \centering
         \includegraphics[width=\textwidth]{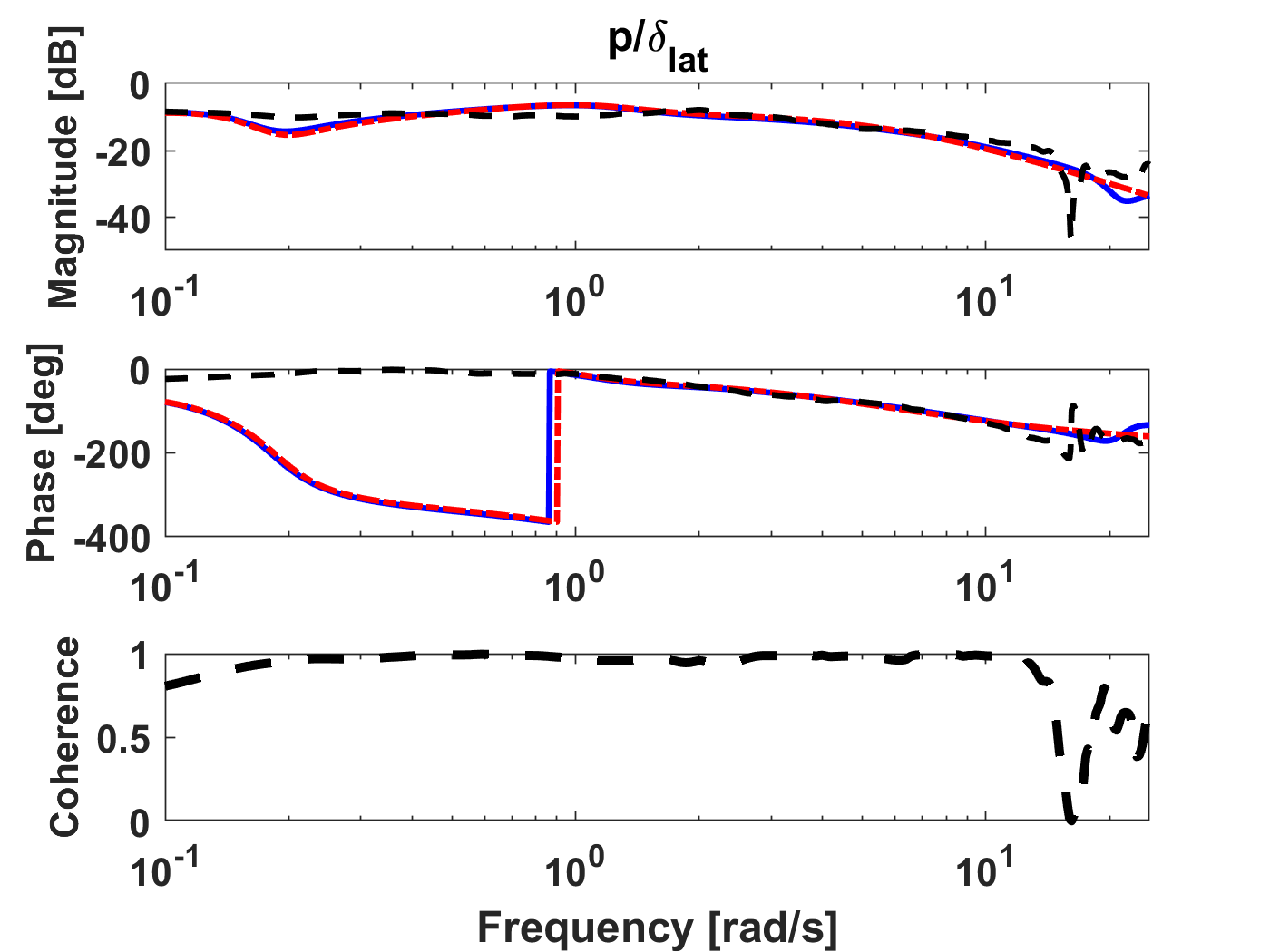}
         \caption{}
         \label{fig:on_axis_open_loop1_a}
     \end{subfigure}
     \hfill
     \begin{subfigure}[b]{0.56\textwidth}
         \centering
         \includegraphics[width=\textwidth]{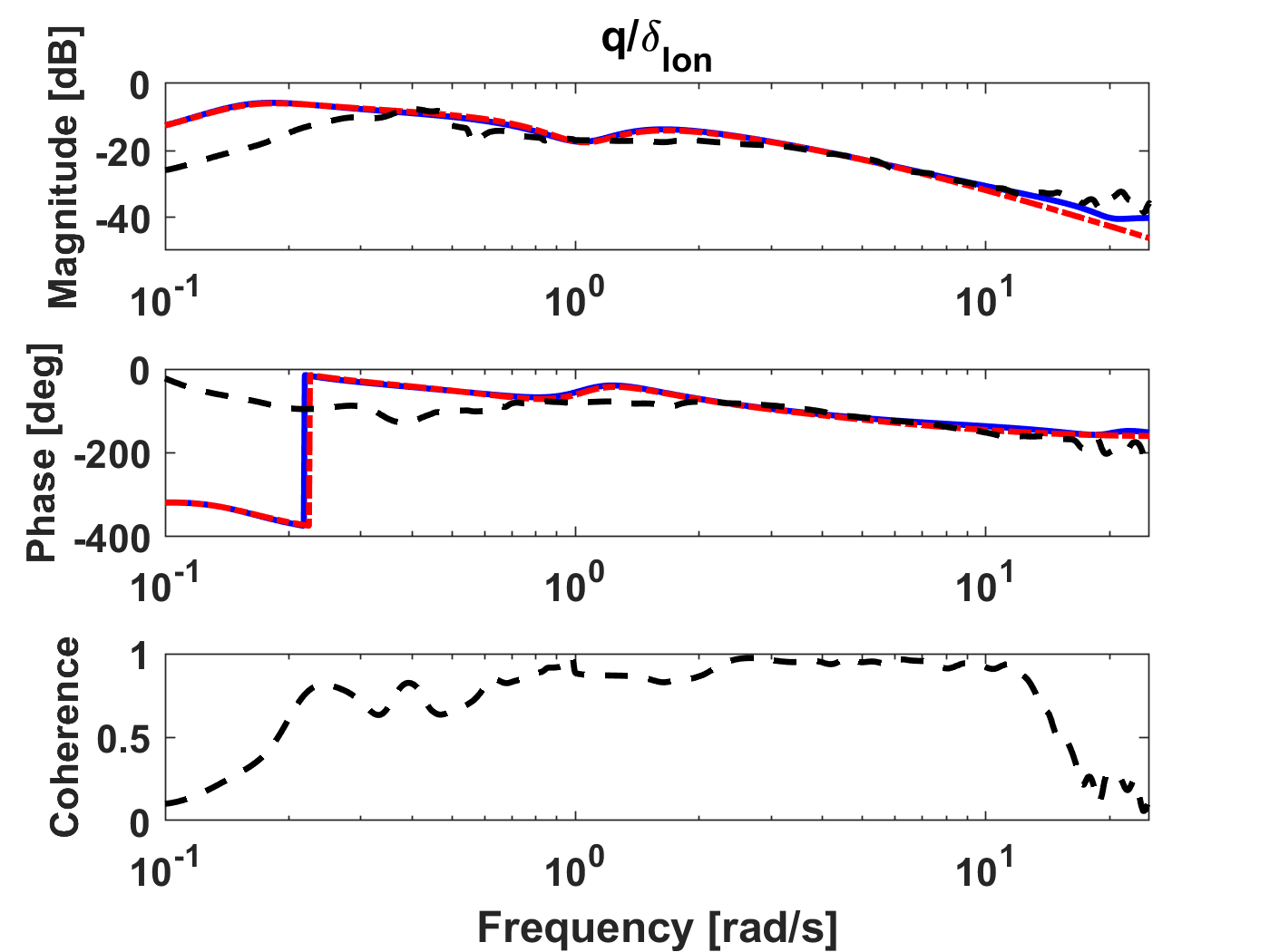}
         \caption{}
         \label{fig:on_axis_open_loop1_b}
     \end{subfigure}
     \hfill
     \begin{subfigure}[b]{0.56\textwidth}
         \centering
         \includegraphics[width=\textwidth]{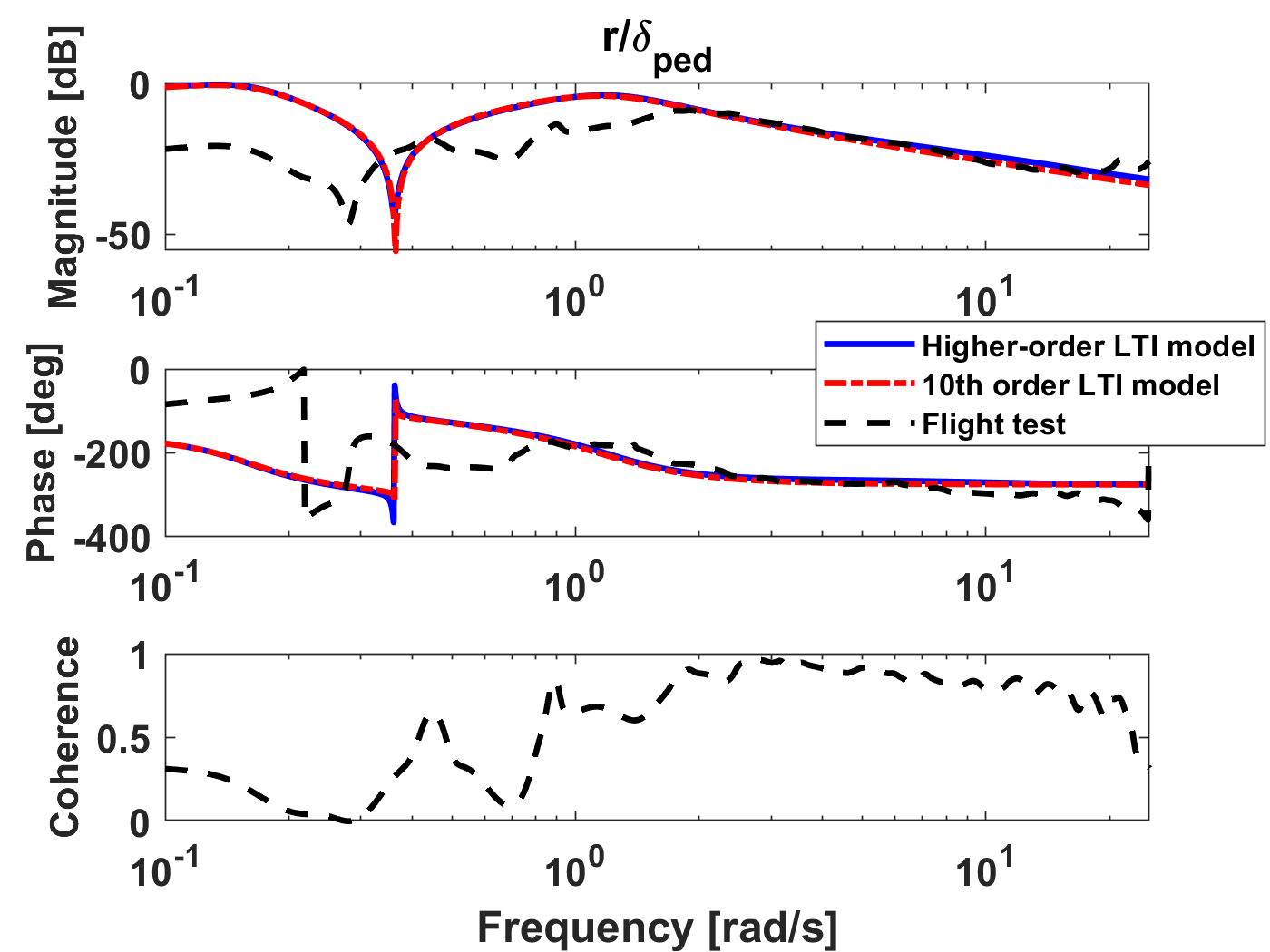}
         \caption{}
         \label{fig:on_axis_open_loop1_c}
     \end{subfigure}
        \caption{On-axis frequency domain response comparison: body states.}
        \label{fig:on_axis_open_loop1}
\end{figure}

\begin{figure}
     \centering
     \begin{subfigure}[b]{0.56\textwidth}
         \centering
         \includegraphics[width=\textwidth]{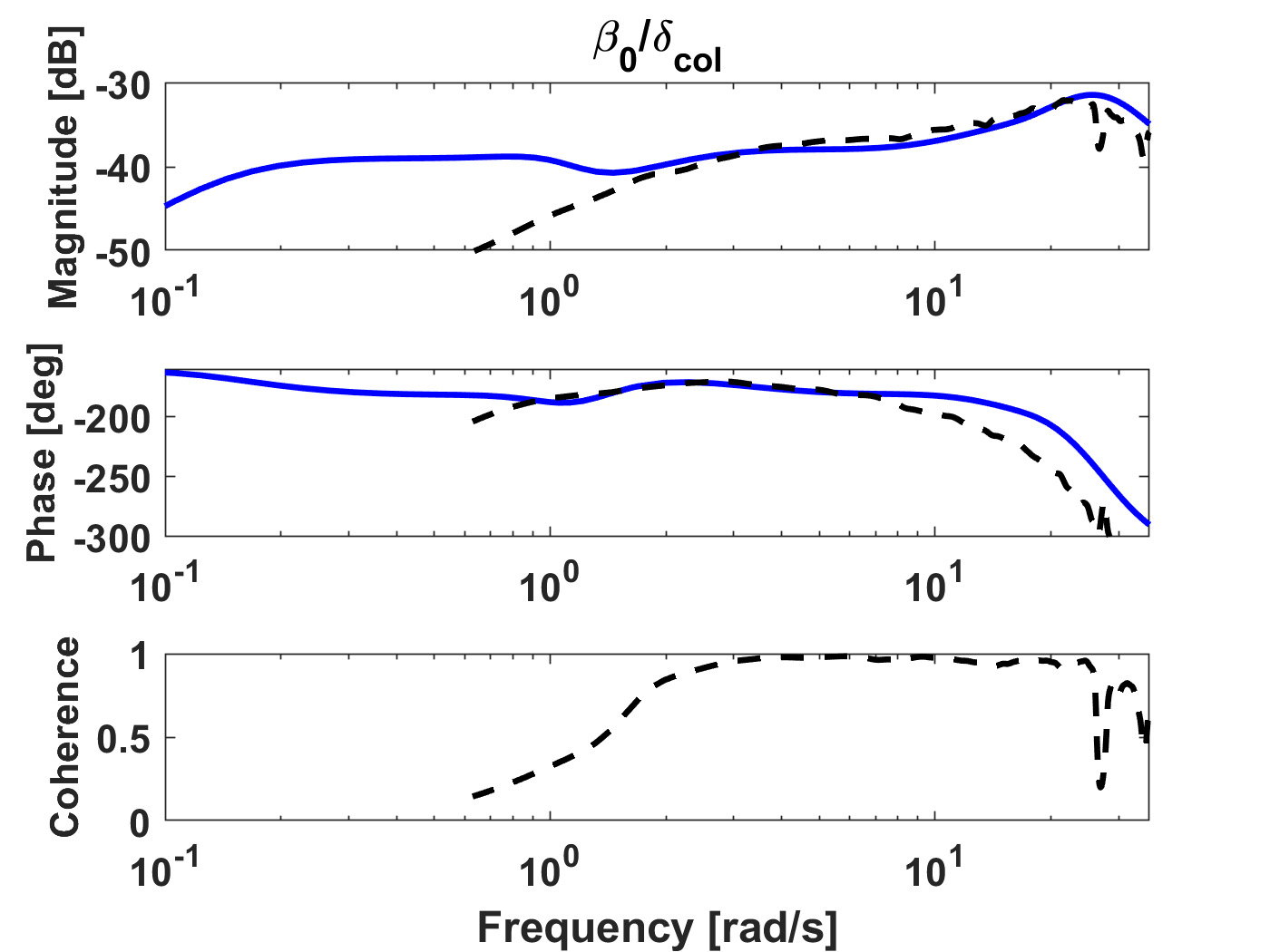}
         \caption{}
         \label{fig:on_axis_open_loop2_a}
     \end{subfigure}
     \hfill
     \begin{subfigure}[b]{0.56\textwidth}
         \centering
         \includegraphics[width=\textwidth]{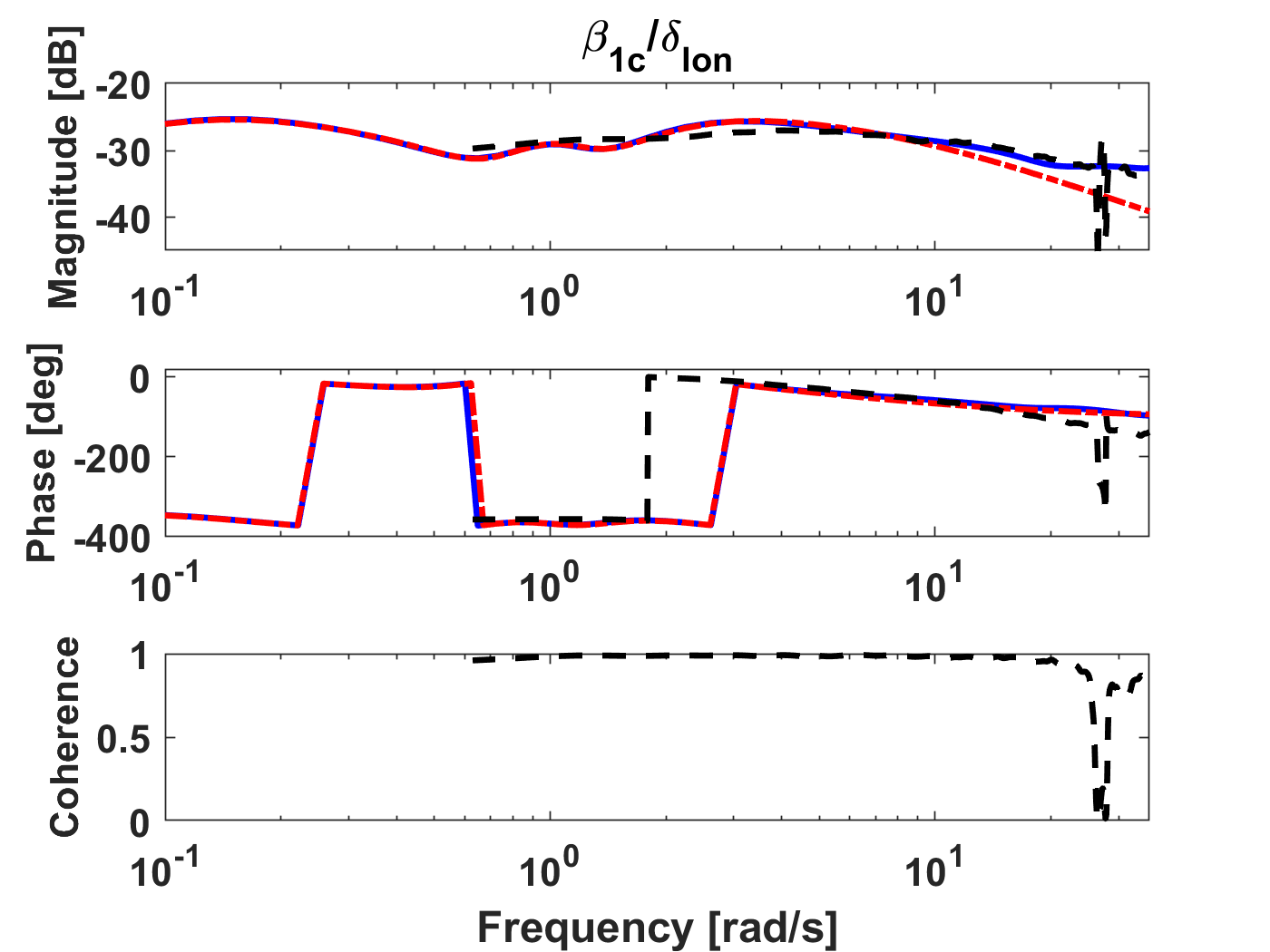}
         \caption{}
         \label{fig:on_axis_open_loop2_b}
     \end{subfigure}
     \hfill
     \begin{subfigure}[b]{0.56\textwidth}
         \centering
         \includegraphics[width=\textwidth]{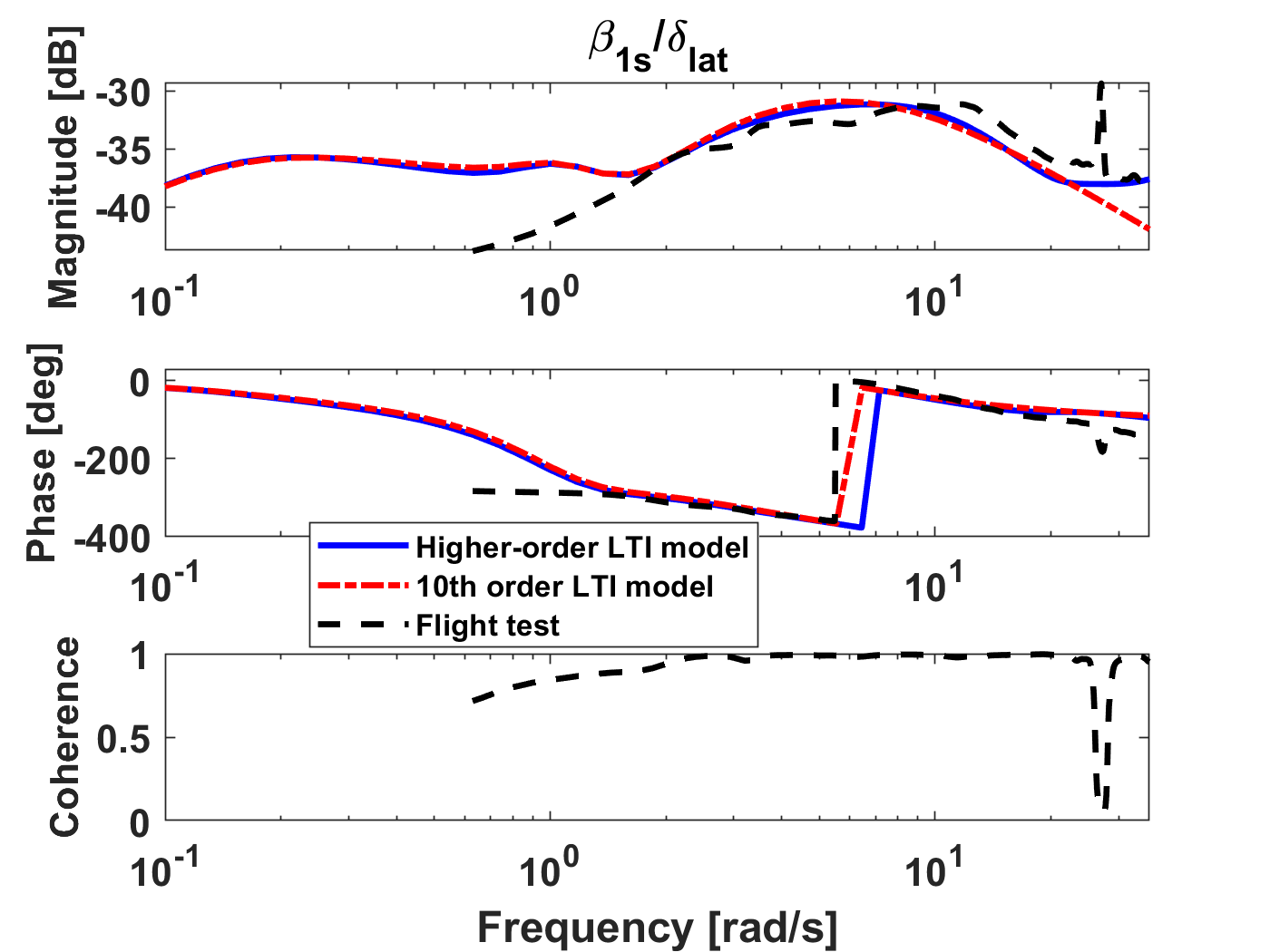}
         \caption{}
         \label{fig:on_axis_open_loop2_c}
     \end{subfigure}
        \caption{On-axis frequency domain response comparison: rotor states.}
        \label{fig:on_axis_open_loop2}
\end{figure}

The higher-order and $10^{\mathrm{th}}$ order LTI models are validated using frequency domain flight data from a JUH-60A Black Hawk at 80 knots, provided by the U.S. Army Aviation Development Directorate (ADD). The validation of both models is conducted at 80 knots, as it is the only flight condition for which frequency domain flight data is available. This data was collected for the Army/NASA Rotorcraft Aircrew Systems Concepts Airborne Laboratory (RASCAL) program. The validation step involves comparing the on-axis, open-loop frequency responses for body angular velocities ($p$, $q$, $r$), coning angle ($\beta_{0}$), longitudinal flapping ($\beta_{\mathrm{1c}}$), and lateral flapping ($\beta_{\mathrm{1s}}$). Figures~\ref{fig:on_axis_open_loop1} and~\ref{fig:on_axis_open_loop2} suggest a good match between the higher-order LTI model frequency response, the $10^{\mathrm{th}}$ order LTI model frequency response, and the flight data, particularly for high coherence values (>$0.6$). For low coherence values (<$0.6$), no definitive conclusions can be drawn given the unreliability of the flight data. It is important to emphasize that the results in Figs.~\ref{fig:on_axis_open_loop1} and~\ref{fig:on_axis_open_loop2} are obtained using a higher-order LTI model, which is extracted at a forward flight speed of 80 knots using the procedure described in Section II. The $10^{\mathrm{th}}$ order LTI model is derived from the higher-order LTI model using the model order reduction technique outlined in Section III.

A metric based on the weighted sum of magnitude and phase squared errors can be used to quantify the fidelity of a simulation model frequency response. First introduced by Hogkinson~\cite{Hodgkinson} and later improved by Tischler~\cite{ref60}, the metric is

\begin{equation}
{J}= \frac{20}{{n_{\omega}}}\sum_{\omega_1}^{\omega_{n_{\omega}}} W_{\gamma} \left[ W_{g} \left( \left| \hat{T}_{c} \right| - \left| \hat{T} \right| \right)^{2} + W_{p} \left( \angle \hat{T}_{c} - \angle \hat{T} \right)^{2} \right]
\label{eq:46_chams2}
\end{equation}

\noindent where $|.|$ and $\angle$ represent the magnitude in decibels and the phase in degrees, respectively, at each frequency $\omega$, ${n_{\omega}}$ is the number of frequency points, and $\omega_{1}$ and $\omega_{n_{\omega}}$ are the starting and ending frequencies of the fidelity assessment, respectively.
\noindent $W_{g}$ and $W_{p}$ are relative weighting coefficients for the magnitude and phase squared errors, respectively. It is a standard practice to set $W_{g}=1.0$ and $W_{p}=0.01745$~\cite{ref60}. This choice of weighting coefficients ensures that 1 dB of magnitude error is equivalent to $7.57^{\circ}$
of phase error.
$W_{\gamma}$ is a weighting coefficient biased towards data with high coherence values. This bias prioritizes reliable flight test data and disregards unreliable data. The level of reliability of the flight test data at each frequency, $\omega$, can be determined using the coherence function, $\gamma^{2}_{xy}$. 

\begin{equation} W_{\gamma}(\omega)=[1.58(1-e^{-\gamma^{2}_{xy}})]^{2}\label{eq:1} \end{equation} 

\noindent The cost metric is employed to quantify the fidelity of both the higher order and the $10^{\mathrm{th}}$ order LTI models by comparing them with the flight test data in the frequency domain.

\begin{table}[ht]
\centering
\caption[Higher Order LTI Model Fidelity in Frequency domain]{Quantification of higher order and $10^{\mathrm{th}}$ order LTI error responses using frequency domain metric}
\label{tbl:sampleTbl1}
\resizebox{0.8\textwidth}{!}{%
\begin{tabular}{llll}

     &  & \hspace{1.7cm} Cost  & \\ \hline
Frequency response & Frequency range (rad/s) & Higher order & $\mathrm{10^{th}}$order\\ \hline
$p/{\delta_{\mathrm{lat}}}$ & 0.6-20  & 66.5 & 77.3 \\
$q/{\delta_{\mathrm{lon}}}$ & 0.6-20  & 32.2 & 62.5 \\
$r/{\delta_{\mathrm{ped}}}$ & 0.6-20  & 247.5 & 240.9 \\ 
$\beta_{\mathrm{0}}/{\delta_{\mathrm{col}}}$ & 0.6-20  & 45.6 & -- \\
$\beta_{\mathrm{1c}}/{\delta_{\mathrm{lon}}}$ & 0.6-20  & 39.4 & 69.5\\
$\beta_{\mathrm{1s}}/{\delta_{\mathrm{lat}}}$ & 0.6-20  & 119.4 & 135.2\\
\hline
     &  & Average: 91.8 & Average: 117.1 \\ \cline{3-4}
\end{tabular}%
}
\end{table}

From Table~\ref{tbl:sampleTbl1}, it can be observed that, across all frequency responses, the higher-order LTI model yields an average cost function value of 91.8, while the $10^{\mathrm{th}}$ order LTI model results in an average cost of 117.1. According to the guideline in~\cite{ref60}, an average cost function value less than or equal to 100 signifies an acceptable level of accuracy for flight dynamics modeling, even if some individual cost functions reach values of 200. 

In comparing the higher-order and $10^{\mathrm{th}}$ order LTI models to flight test results, a higher metric value of approximately $J=240$ was observed in the $\frac{r}{\delta_{\mathrm{ped}}}$ transfer function. This loss of fidelity in the yaw axis has been observed in previous studies aimed at developing a simulation model of the UH-60 aircraft~\cite{ref_turnour, ref_saetti_thesis}. A more accurate model could be obtained by further tuning or increasing the fidelity of the model (for instance, improving the tail rotor model) to accurately capture the important phenomena that affect the vehicle response in the yaw axis. Since the aim of this work is to have a generic UH-60 aircraft, no further model tuning or improvement is performed. Therefore, it can be concluded that the extracted higher order LTI model accurately represents the flight dynamics of a generic UH-60 aircraft. The finding also suggests that the developed nonlinear model is representative of a UH-60 aircraft in terms of the body, coning, longitudinal flapping, and lateral flapping responses. The $10^{\mathrm{th}}$ order LTI model represents a very good approximation of the higher-order LTI model at low to medium frequencies, as observed in Figs.~\ref{fig:on_axis_open_loop1} and~\ref{fig:on_axis_open_loop2}. However, at high frequencies, the $10^{\mathrm{th}}$ order LTI model loses accuracy (see~Figs.~\ref{fig:on_axis_open_loop2_b} and~\ref{fig:on_axis_open_loop2_c}). This is expected since the $10^{\mathrm{th}}$ order LTI model is extracted from the higher order LTI model using a quasi-steady approximation of the fast dynamics. This explains the higher average cost obtained for the $10^{\mathrm{th}}$ order LTI model ($J=117.1$) compared to the higher-order LTI model ($J=91.8$). 



\subsection{Batch simulations}
\subsubsection{Simulation setup}

\begin{figure}[H]
\centering
\includegraphics[width=1.0\textwidth]{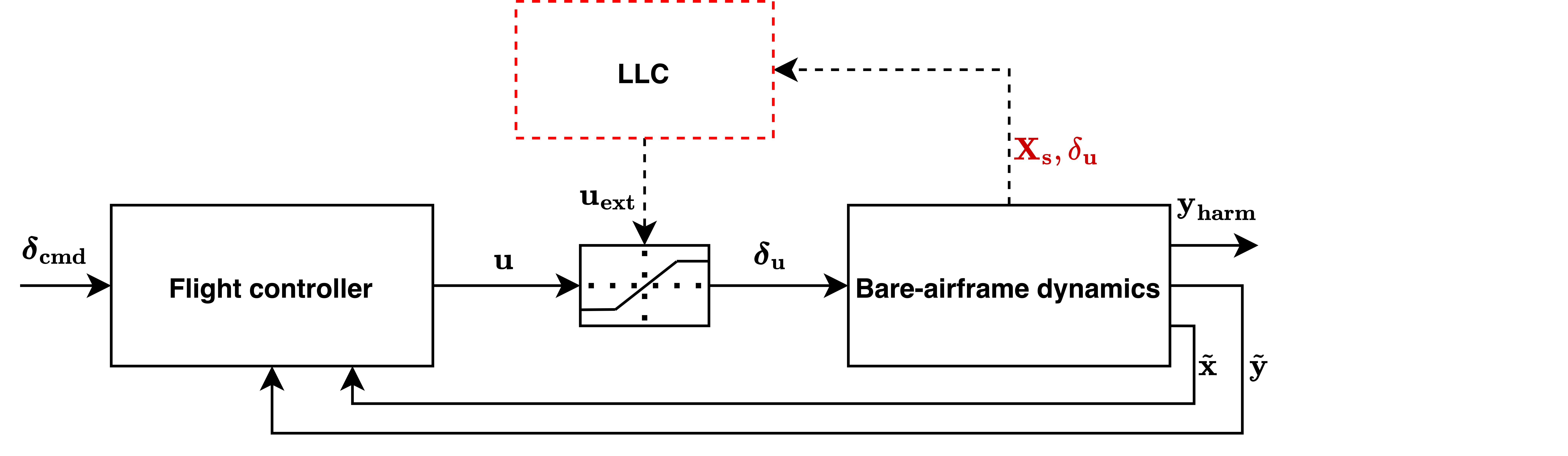}
\caption{LLC scheme via control effector command limiting.}
\label{fig:30_chams}
\end{figure}

\begin{figure}[H]
\centering
\includegraphics[width=1.0\textwidth]{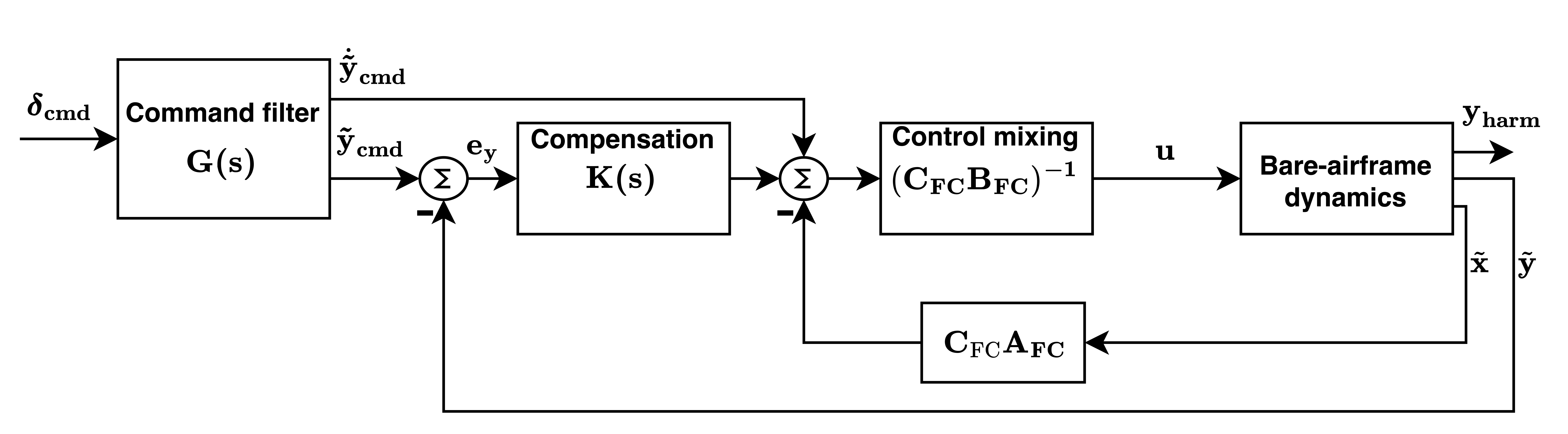}
\caption{Flight controller of Fig.~\ref{fig:30_chams}: dynamic inversion control law (adapted with permission from~\cite{aerospace6030038}).}
\label{fig:DI_controller}
\end{figure}

The LLC scheme computes extremal control inputs, which automatically limit the control effector commands generated by the flight controller, to achieve component load limiting. A block diagram representation of the LLC scheme via control effector command limiting is displayed in Fig.~\ref{fig:30_chams}. In this figure, the bare-airframe model and flight controller correspond to the high-fidelity FLIGHTLAB$\textsuperscript{\textregistered}$ model and a dynamic inversion controller~\cite{aerospace6030038}, respectively. Figure~\ref{fig:DI_controller} provides a more detailed illustration of the flight controller.


The dynamic inversion controller uses linearized aircraft dynamics ($A_{\mathrm{FC}}$, $B_{\mathrm{FC}}$, $C_{\mathrm{FC}}$), scheduled with airspeed, in an inner feedback loop to invert the plant model through feedback linearization. Additionally, feedback compensation through a Proportional-Integral-Derivative (PID) controller is employed, allowing $\tilde{y}$ to track the reference signal $\tilde{y}_{\mathrm{cmd}}$. The transfer function, $G(\mathrm{s})$, within the command filter determines the desired dynamics of the closed-loop or the nature of the desired response type.  One response type is Attitude Command/Attitude Hold (ACAH), correlating pilot stick displacement with vehicle attitude. Another response type is Rate Command/Attitude Hold (RCAH), relating pilot stick displacement to vehicle angular rate. The choice of response type heavily relies on the Usable Cue Environment (UCE)~\cite{ref34}. 

\begin{figure}[H]
\centering
\includegraphics[width=1.0\textwidth]{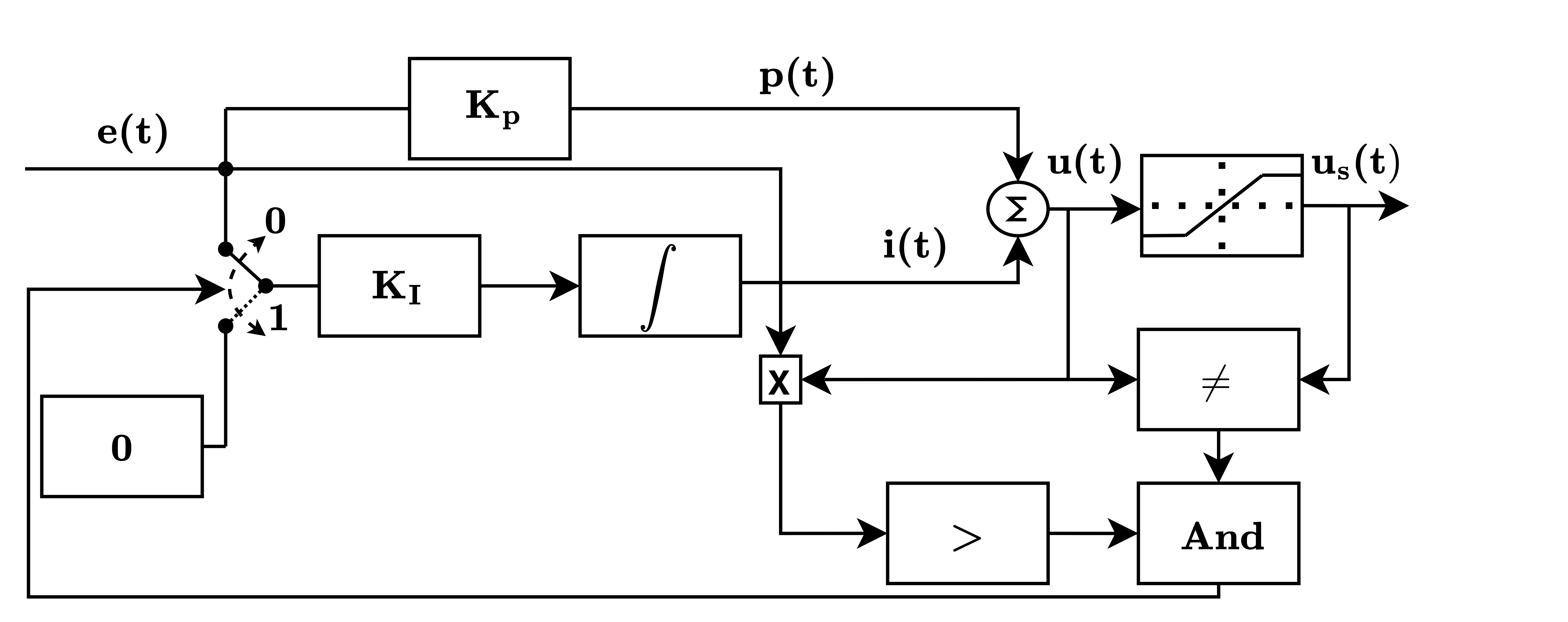}
\caption{Integrator anti-windup scheme (developed using data from~\cite{ref_anti4}).}
\label{fig:anti_windup_scheme}
\end{figure}

The interplay between the flight and LLC controllers can have a detrimental impact on maneuver performance. This is attributed to the conflicting nature of the constraints imposed on both controllers. While the LLC controller adjusts the aircraft's response to maintain the harmonic load within the specified limit by restricting control effector commands, the flight controller aims to ensure the aircraft's response follows the reference trajectory. This dynamic interaction may result in the integrator windup effect~\cite{ref_anti1,ref_anti2,ref_anti3}, a nonlinear phenomenon common in controllers with PID feedback compensation. The phenomenon is characterized by significant increase in the integral of the error between $\tilde{y}$ and $\tilde{y}_{\mathrm{cmd}}$, a direct consequence of input saturation, which leads the system to experience large overshoot and a poor transient response. To remediate this issue, an anti-windup scheme~\cite{ref_anti2,ref_anti3,ref_anti4,ref_anti5} presented in Fig.~\ref{fig:anti_windup_scheme} is added to the flight controller.


To test the performance of the LLC scheme in batch simulations, two pitch maneuvers were conducted. The first is an attitude command maneuver where the flight controller is set to give an Attitude Command/Attitude Hold (ACAH) response. The second maneuver is a rate command maneuver where the flight controller is set to give a Rate Command/Attitude Hold (RCAH) response.

\subsubsection{Attitude command maneuver}

\begin{figure}[H]
\centering
\includegraphics[width=1.0\textwidth]{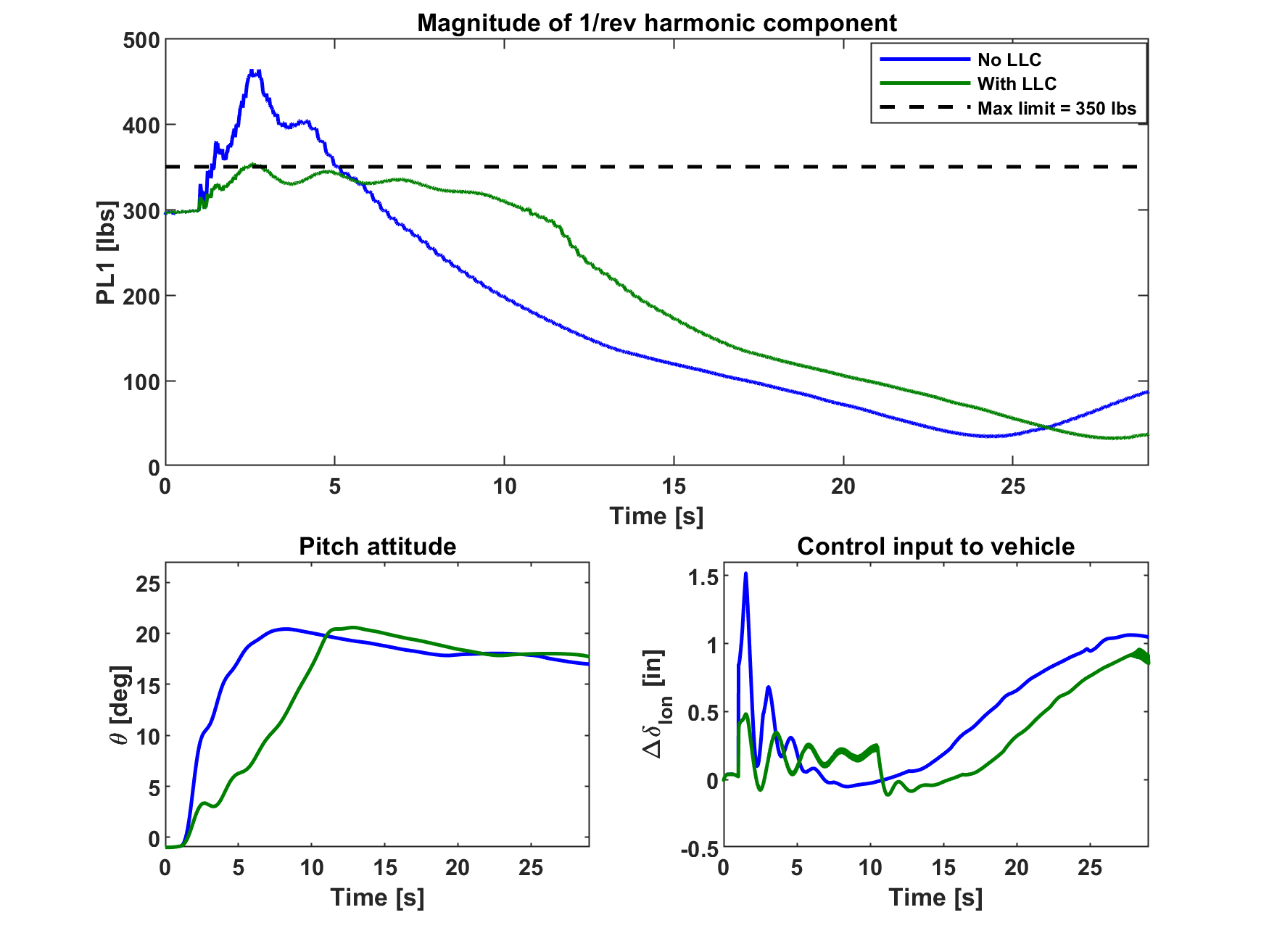}
\caption{Attitude command maneuver.}
\label{fig:perf_1}
\end{figure}

For the attitude command maneuver, the longitudinal cyclic stick input from the pilot is taken to be the desired pitch attitude command input. The pilot applies a step input, commanding the vehicle to reach and maintain an attitude of approximately 20 degrees until the step input is removed. Simulation results for the reference blade root 1/rev pitch link load magnitude, vehicle pitch attitude, and longitudinal cyclic control input, both with and without LLC, are depicted in Fig.~\ref{fig:perf_1}. It is evident from the figure that the 1/rev magnitude of the pitch link load of the nonlinear model mostly stays within the specified 350 lbs limit with LLC. However, there is a slight load exceedance at around $t=3$ seconds, attributable to two factors. First, despite the reduced-order LTI model used in formulating the proposed LLC scheme can be used for real-time component load estimation, the estimated loads may require correction due to errors arising from LTI model approximations and nonlinearities~\cite{ref28, ref29, ref30}. Consequently, the on-board linear model may experience a loss of fidelity under substantial inputs. Second, the MPC algorithm employs a linear approximation to represent the 1/rev harmonic load of the nonlinear model. Hence, some limit exceedance is expected. Nevertheless, the LLC scheme demonstrates effective performance in maintaining the pitch link load within the specified limit.

The impact of the load limiting control strategy on helicopter maneuver performance is demonstrated by comparing the achieved pitch attitude in cases with and without LLC. Figure~\ref{fig:perf_1} illustrates that, with the LLC scheme, the desired pitch attitude is attained. The trade-off between load limiting and maneuver performance is evident in the time required to reach the desired attitude. Maneuver aggressiveness is constrained to control the resulting 1/rev magnitude of pitch link load, subsequently leading to a delayed attainment of the desired pitch attitude.

It is essential to highlight the integration of an integrator anti-windup scheme into the flight controller, as this addition enables the LLC scheme to achieve effective component load limiting by modifying the control effector commands (see $\Delta \delta_{\mathrm{lon}}$ in Fig.~\ref{fig:perf_1}) while mitigating attitude overshoot. Another noteworthy observation is that the LLC scheme introduces some chattering near the load limit, which, in turn, affects the vehicle pitch attitude response. This chattering phenomenon is a recognized issue in envelope protection systems that restrict control inputs near the limit~\cite{ref_chat1, ref_chat2}. Near the limit, control limiting algorithms can exhibit behavior resembling a nonlinear feedback controller with high gain, leading to such oscillations. To address this challenge,~\cite{ref_chat2, ref_chat3} recommend implementing smoothing algorithms or logics at the limit.

\subsubsection{Rate command maneuver}

\begin{figure}[H]
\centering
\includegraphics[width=1\textwidth]{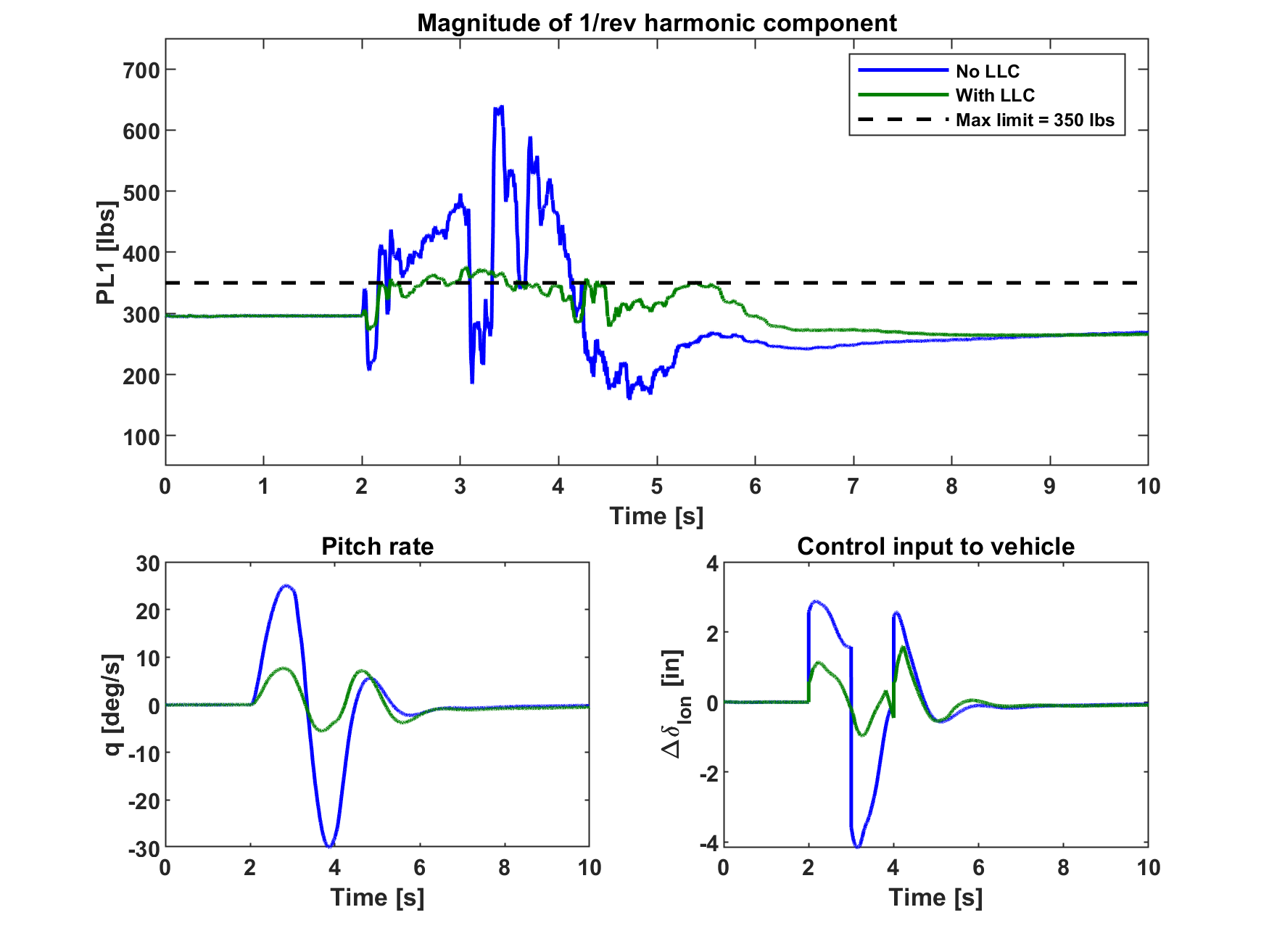}
\caption{Rate command maneuver.}
\label{fig:perf_2}
\end{figure}

In the rate command maneuver, the pilot longitudinal cyclic stick input directly maps to the desired pitch rate command. Figure~\ref{fig:perf_2} illustrates simulation results from a 2-second pilot doublet input, presenting the reference blade root 1/rev pitch link load magnitude, vehicle pitch attitude, and longitudinal cyclic control input—with and without LLC. While the LLC scheme effectively maintains the 1/rev pitch link load mostly within the 350 lb limit, it does come with a trade-off of significantly diminished pitch rate maneuver performance. This is attributed to the selected load limit being close to the trim load and the crucial role of 1/rev control inputs, transmitted through the pitch link, in enabling aircraft maneuvering. Future work should explore a less conservative load limit and consider restricting higher harmonic pitch link loads (e.g., 2/rev, 3/rev, 4/rev, etc.) instead of the 1/rev load to mitigate the impact of the scheme on maneuver performance.

\subsection{Pilot-in-the-loop simulations}

\subsubsection{Experiment setup}

\begin{figure}[h!]
\centering
\includegraphics[width=0.7\textwidth]{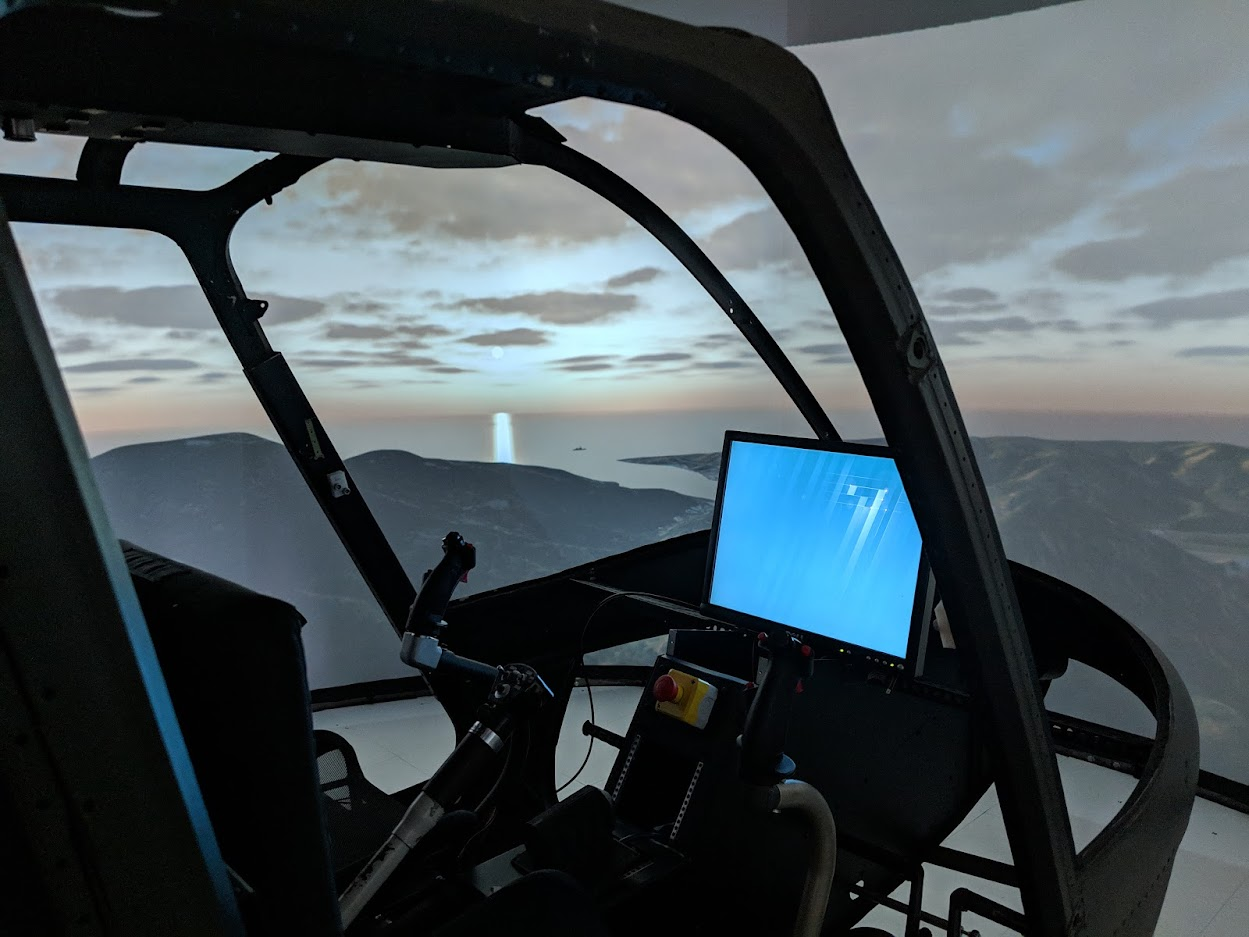}
\caption{Rotorcraft flight simulator.}
  \label{fig1:sim}
\end{figure}

In this section, the performance of the LLC scheme is assessed through piloted simulations utilizing a fixed-base OH-58D cockpit as shown in Fig.~\ref{fig1:sim}. The simulator imagery is displayed on a screen with a 16-feet diameter and 270-degree field of view. The simulator is equipped with a mechanically linked control system with no advanced features (i.e., no ACAH/RCAH modes). The LLC scheme is integrated with a cueing algorithm and the resulting architecture is implemented within the simulator. Piloted flight simulation experiments are carried out to assess the effectiveness of the cue in limiting control input aggressiveness.

Significant research efforts have been devoted to advancing aural, visual, and tactile cueing mechanisms to facilitate envelope protection and augment pilot perception in visually challenging environments~\cite{ref_cueing1, ref_cueing2, ref_cueing3, ref_cueing4}. Each of these cueing methodologies presents distinct advantages and is most suitable for specific applications. In this study, a visual cueing architecture is favored over tactile or aural cueing architectures. This preference is grounded in two primary considerations.~First, given that simulations take place in good visual environments and the cue conveys only a single piece of information, opting for a visual cue minimizes intrusion when compared to tactile and aural cues.~Second, a visual cue provides the pilot with a straightforward means of quantifying the available control margin throughout the flight. Nonetheless, forthcoming studies will concentrate on evaluating the performance of tactile and aural cues in comparison to a visual cue. It is clear, however, that in a degraded visual environment, tactile and aural cues would represent better options than a visual cue.

\begin{figure}[H] \begin{center}
     \begin{subfigure}[b]{0.3\textwidth}
         \centering
         \includegraphics[width=\textwidth]{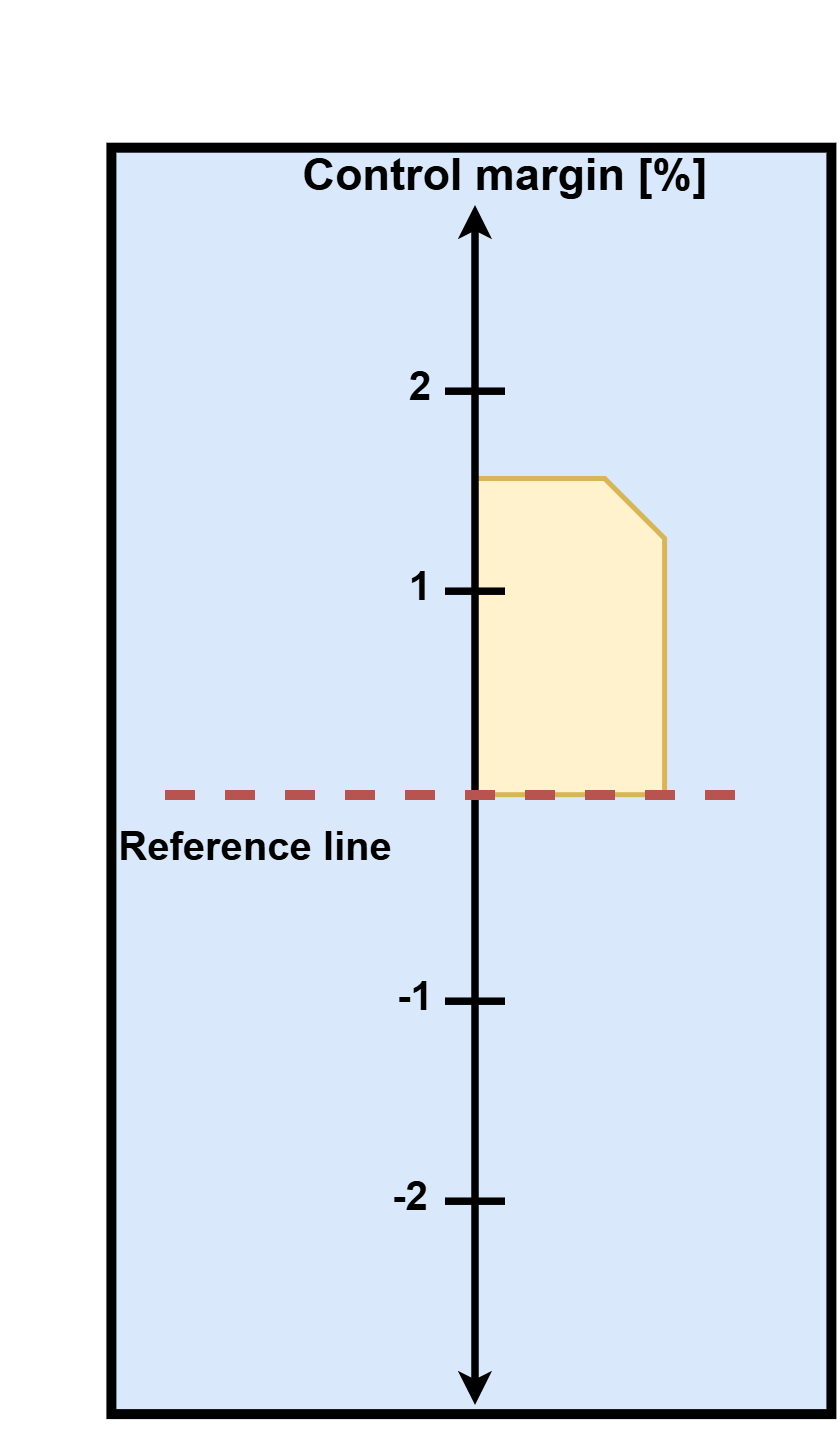}
         \caption{Nominal case with control margin}
         \label{fig1: chap_17a}
     \end{subfigure}
     \hfill
     \begin{subfigure}[b]{0.3\textwidth}
         \centering
         \includegraphics[width=\textwidth]{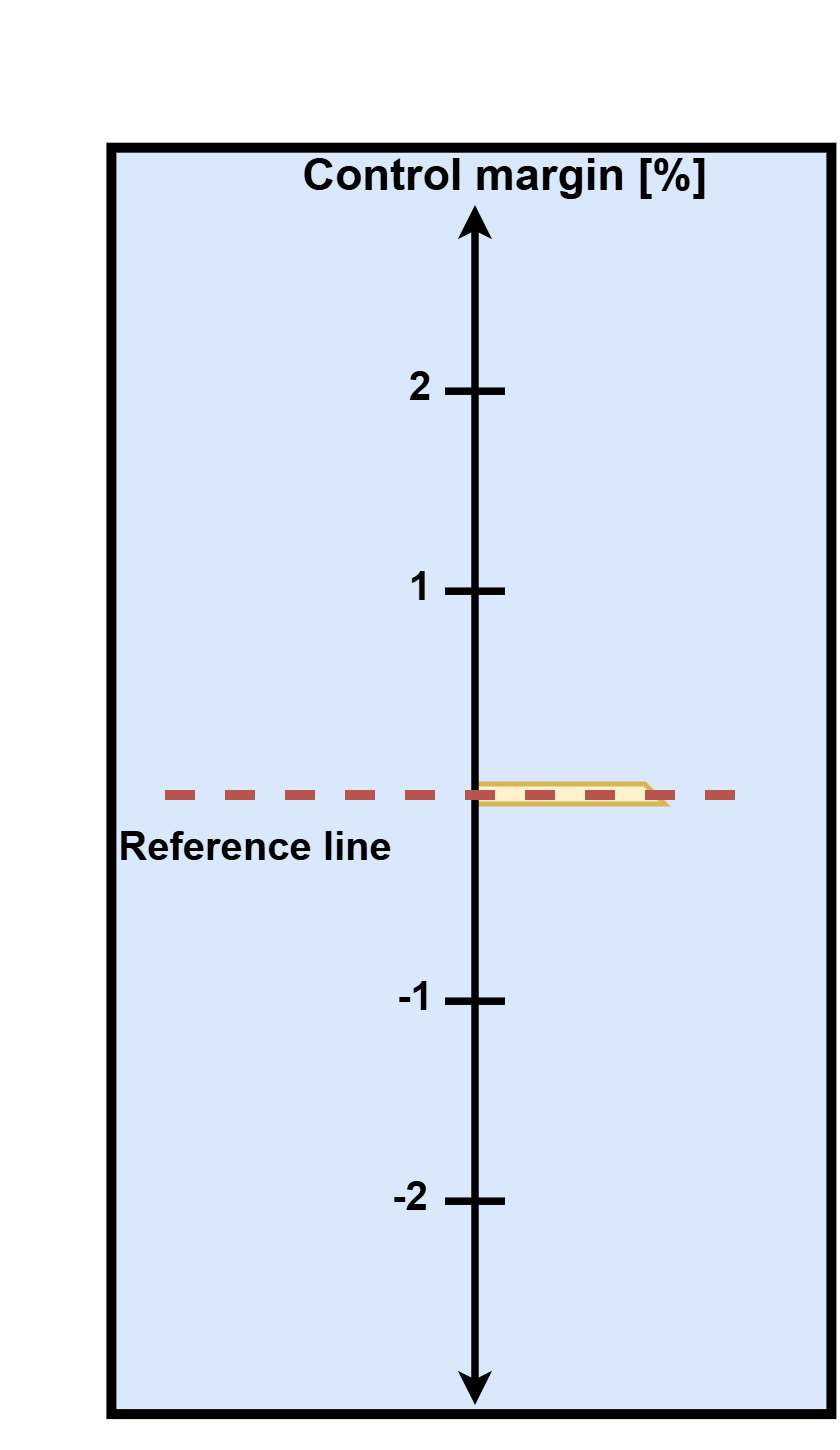}
         \caption{Case with zero control margin}
         \label{fig1: chap_17b}
     \end{subfigure}
       \hfill
     \begin{subfigure}[b]{0.3\textwidth}
         \centering
         \includegraphics[width=\textwidth]{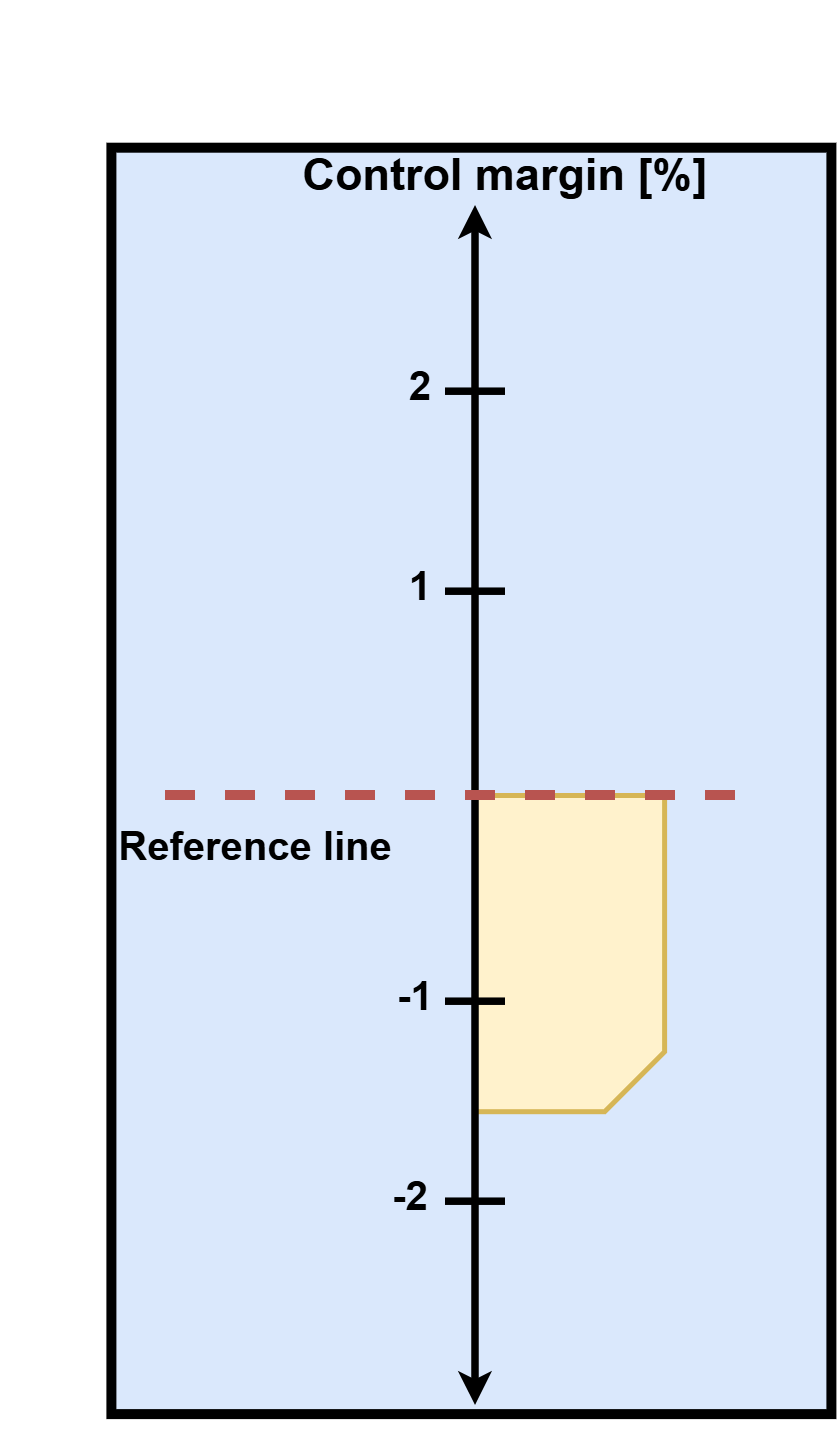}
         \caption{Case with negative control margin}
         \label{fig1: chap_17c}
     \end{subfigure}
        \caption{2D visual cue.}
        \label{fig1: chap_17}
\end{center}\end{figure}

The cueing algorithm was developed with the assistance of a former U.S. Army Black Hawk pilot who conducted the simulations presented. This algorithm takes the control margin (Eq.~\ref{eq:CM}) as input and generates a visual cue. The conversion from control margin to visual cue involves rescaling the control margin using a constant gain and transforming the obtained quantity into a 2D graphic. This rescaling gain is tuned based on the pilot's preference. Notably, no filtering was applied to the control margin before it was used by the cueing algorithm. This decision was made to gain an initial understanding of the dynamic nature of the cue and to assess the pilot's ability to follow it. It is expected that no filtering would be needed for a 1/rev load. However, for higher harmonic loads, for instance, 2/rev, 3/rev, 4/rev, etc., some level of filtering would be needed. The 2D visual cue emanating from the cueing algorithm is shown in Fig.~\ref{fig1: chap_17}.

\pagebreak

Figure~\ref{fig1: chap_17} shows that one side of the cue aligns with a fixed reference line, while the other side is free to move vertically, causing the cue's height to change over time. The width of the cue does not change with time. At each instant, the cue enables the pilot to gauge the amount of control deflection required to bring the harmonic load from its current value to its limit. When the cue has a height of zero (Fig.~\ref{fig1: chap_17b}), the control margin is zero, implying that the harmonic load (i.e., 1/rev load) has reached its limit. If the cue is below the reference line (Fig.~\ref{fig1: chap_17c}), it indicates a negative control margin, signifying load exceedance, where the harmonic load surpassed the maximum load value. Conversely, a positive control margin, illustrated by the cue being above the reference line (Fig.~\ref{fig1: chap_17a}), suggests that the load limit has not been exceeded.




Real-time piloted simulation evaluations were conducted to assess the effectiveness of the visual cue in limiting control input aggressiveness for component load limiting. Two control input profiles were considered at a forward flight speed of 120 knots. For both control inputs, the LLC scheme was tuned to limit the load to 350 lbs.

\subsubsection{Doublet input}

In this first experiment, the pilot is instructed to perform a pitch doublet control input of $20\%$ magnitude with the cue both off and on. In the cue-off condition, the pilot performs the task without considering the cue, while in the cue-on condition, the pilot incorporates the cue into the task. Throughout the experiment, the pilot maintains both the collective stick and pedal at their trim values, which is crucial since the LLC scheme operates solely on the longitudinal cyclic channel.

\begin{figure}[H]
\centering
\begin{subfigure}[b]{0.69\textwidth}
    \centering
    \includegraphics[width=\textwidth]{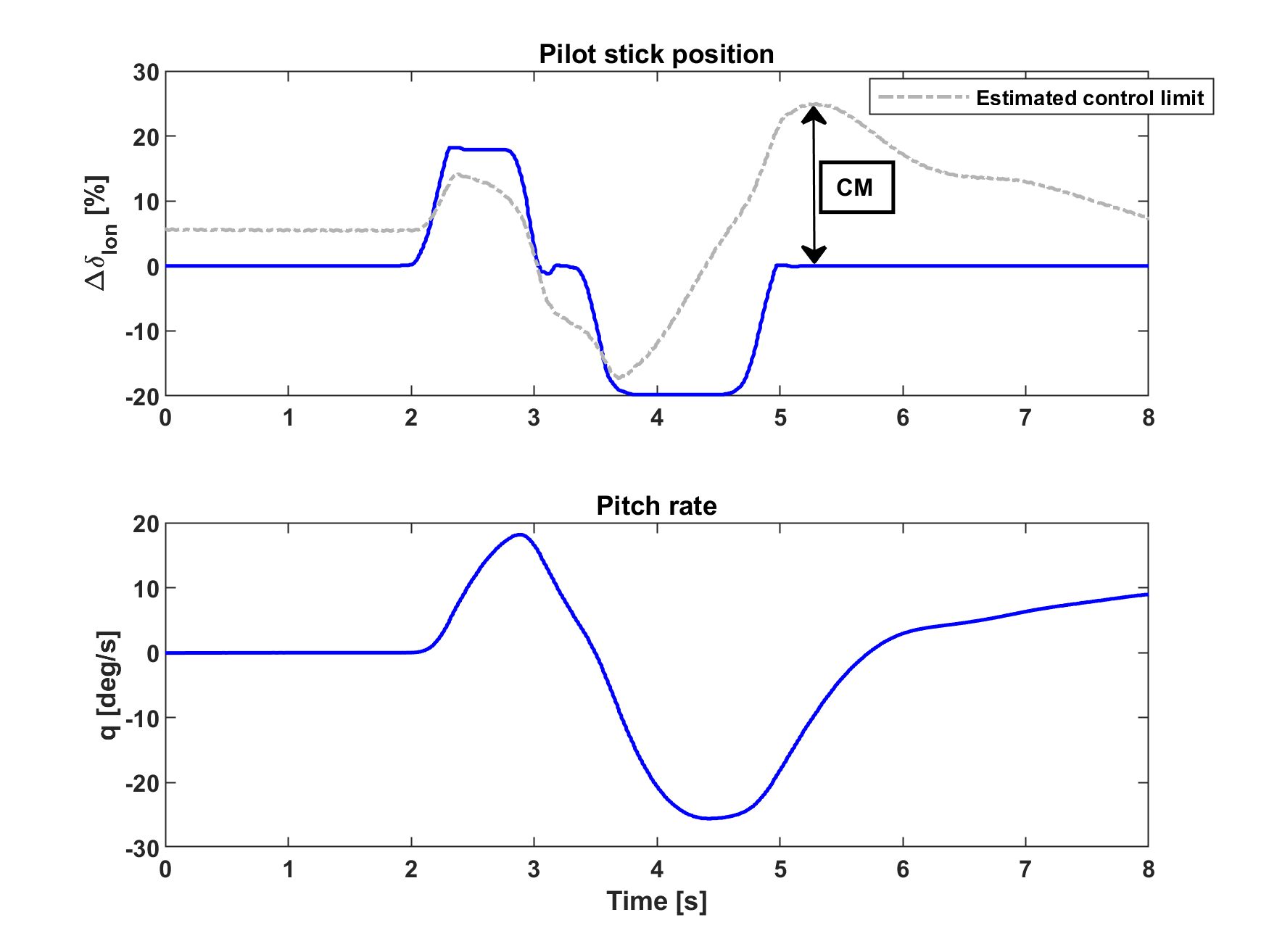}
    \caption{Cue off}
    \label{fig:left_a}
\end{subfigure}
\hspace{-0.065\textwidth}
\begin{subfigure}[b]{0.69\textwidth}
    \centering
    \includegraphics[width=\textwidth]{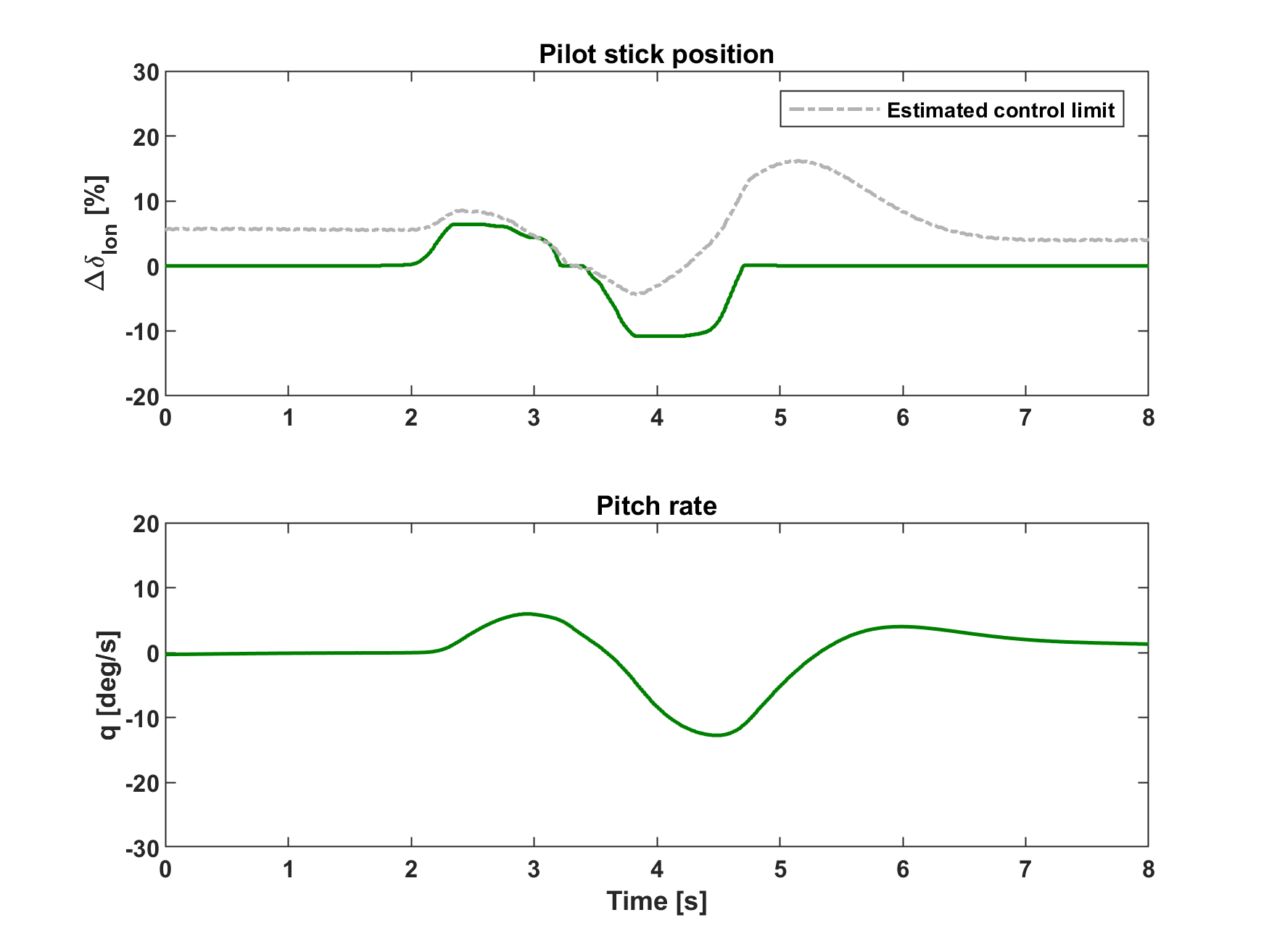}
    \caption{Cue on}
    \label{fig:right_b}
\end{subfigure}

\caption{First experiment: pilot stick position and vehicle response.}
\label{fig1:flight_test1}
\end{figure}

\begin{figure}[H]
\centering
\includegraphics[width=0.85\textwidth]{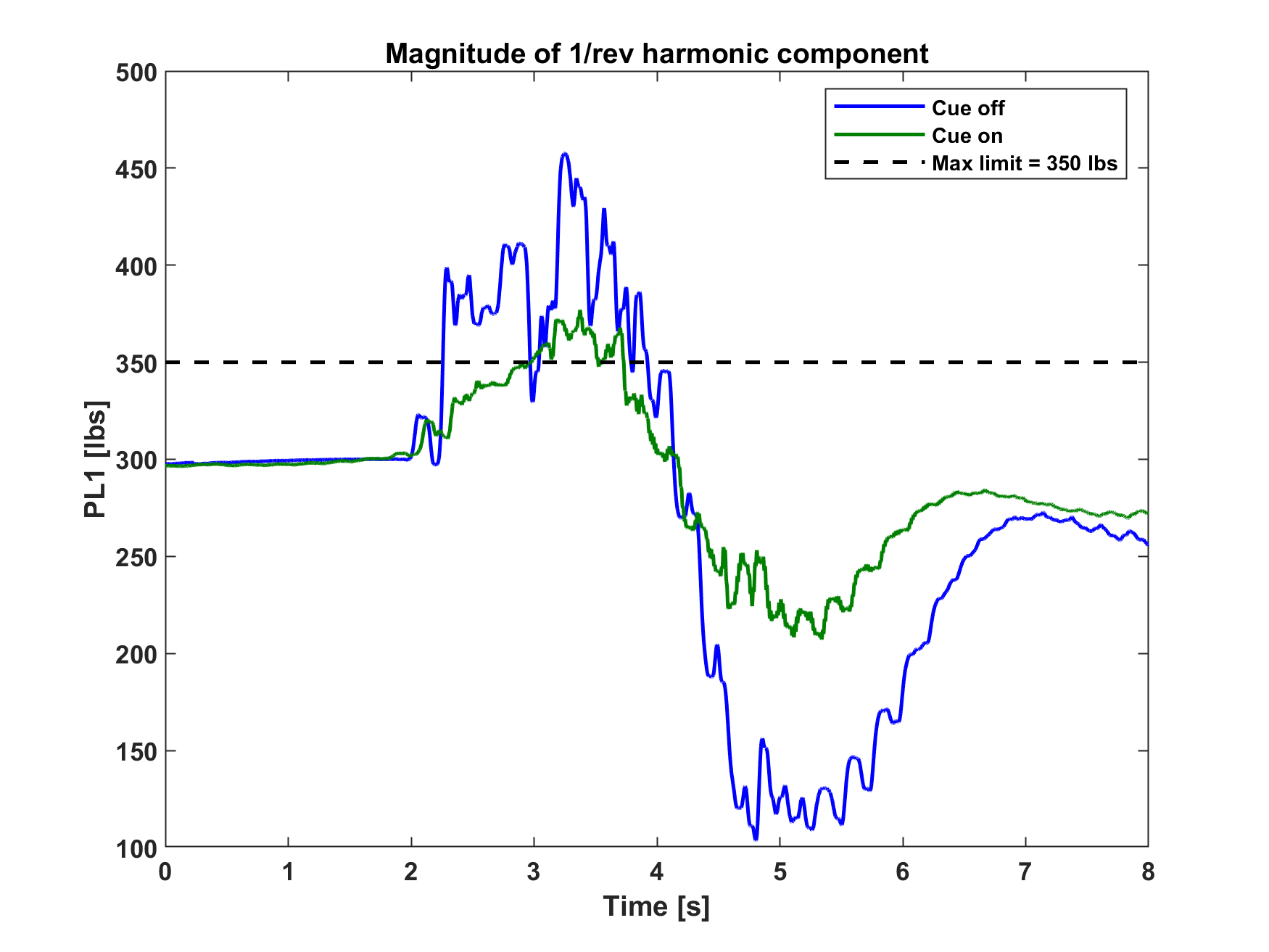}
\caption{First experiment: 1/rev pitch link load.}
  \label{fig1:flight_tes2}
\end{figure}

Figures~\ref{fig1:flight_test1} and~\ref{fig1:flight_tes2} present results from the experiment. Figure~\ref{fig1:flight_test1} illustrates both the pilot stick position, overlaid with the extremal control position, and the vehicle pitch rate. For each simulation, the predicted control margin is represented graphically by the difference between the extremal control position and the pilot stick position. At any given time, the control margin is negative if the pilot stick position exceeds the extremal control position, zero if it matches the extremal control position, and positive if it is below the extremal control position. In the cue-off case, the control margin starts as positive and constant from the beginning of the simulation. However, starting at 2.1 seconds, it decreases rapidly due to the increased aggressiveness of the input, eventually reaching negative values indicating load exceedance. In contrast, in the cue-on case, the pilot stick position remains within the control limit, resulting in a non-negative control margin throughout the simulation. Therefore, by following the cue, the pilot is able to increase the control margin in regions where it was negative during the cue-off case due to aggressive control input, thereby aiming to prevent load exceedance. It is essential to note that the computed control limits for the cue-off and cue-on cases are similar before the start of the input but differ during and after the input due to their dependence on aircraft states.~Therefore, the pilot control stick position corresponding to zero control margin may vary between the two cases, as observed in previous cueing studies for flight envelope protection~\cite{Jeram}.
As a result of the limit imposed on the stick position for the cue-on case, the vehicle response is restricted compared to the cue-off case.

As depicted in Fig.~\ref{fig1:flight_tes2}, the 1/rev load exceeds the specified limit of 350 lbs for the cue-off case, while for the cue-on case, the load is constrained to approximately 350 lbs. Importantly, for the purpose of component life extension, slight load exceedance is acceptable since the user-defined max limit is a soft limit rather than a hard limit. A soft limit is more tolerant of limit violation, resulting in increased operational costs rather than an immediate system failure.~Overall, the results from this experiment demonstrate the impact of load limiting on vehicle maneuverability.

\subsubsection{~Pulse input}

\begin{figure}[H]
\centering
\includegraphics[width=1.0\textwidth]{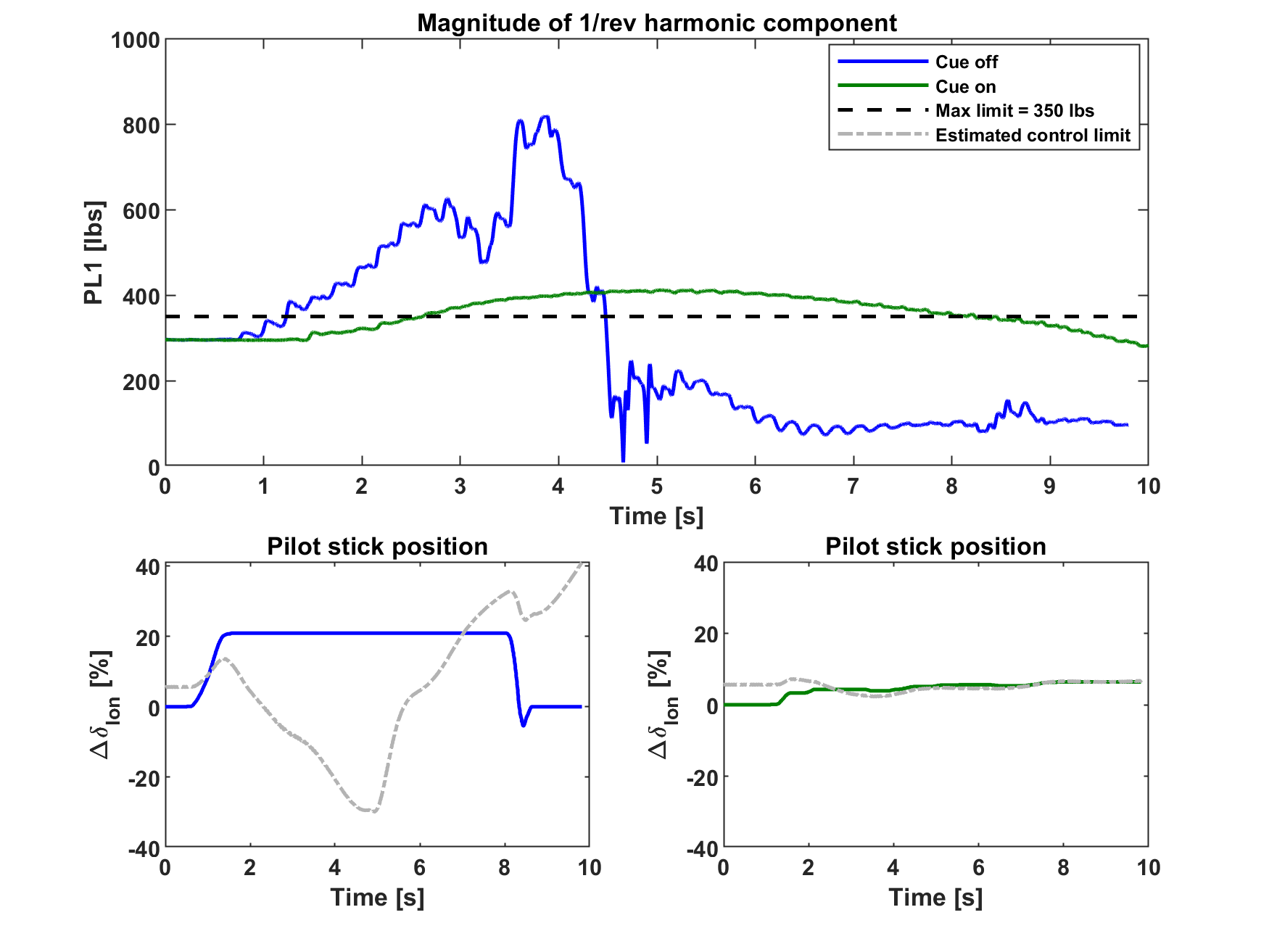}
\caption{Second experiment: pilot stick position and 1/rev pitch link load.}
  \label{fig1:flight_test3}
\end{figure}

\begin{figure}[H]
\centering
\begin{subfigure}[b]{0.85\textwidth}
    \centering
    \includegraphics[width=\textwidth]{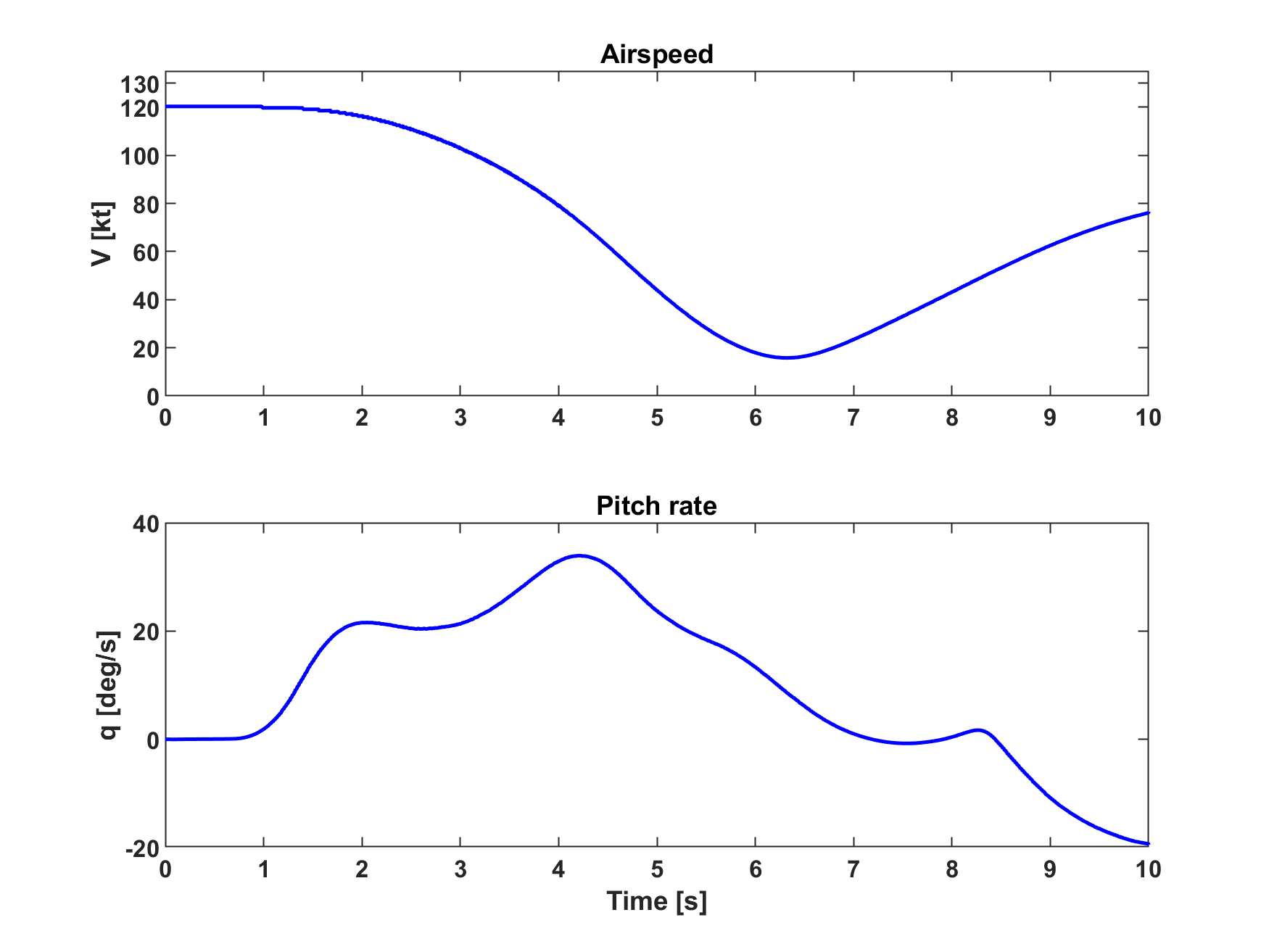}
    \caption{Cue off}
    \label{fig:left_a}
\end{subfigure}
\hspace{-0.065\textwidth}
\begin{subfigure}[b]{0.85\textwidth}
    \centering
    \includegraphics[width=\textwidth]{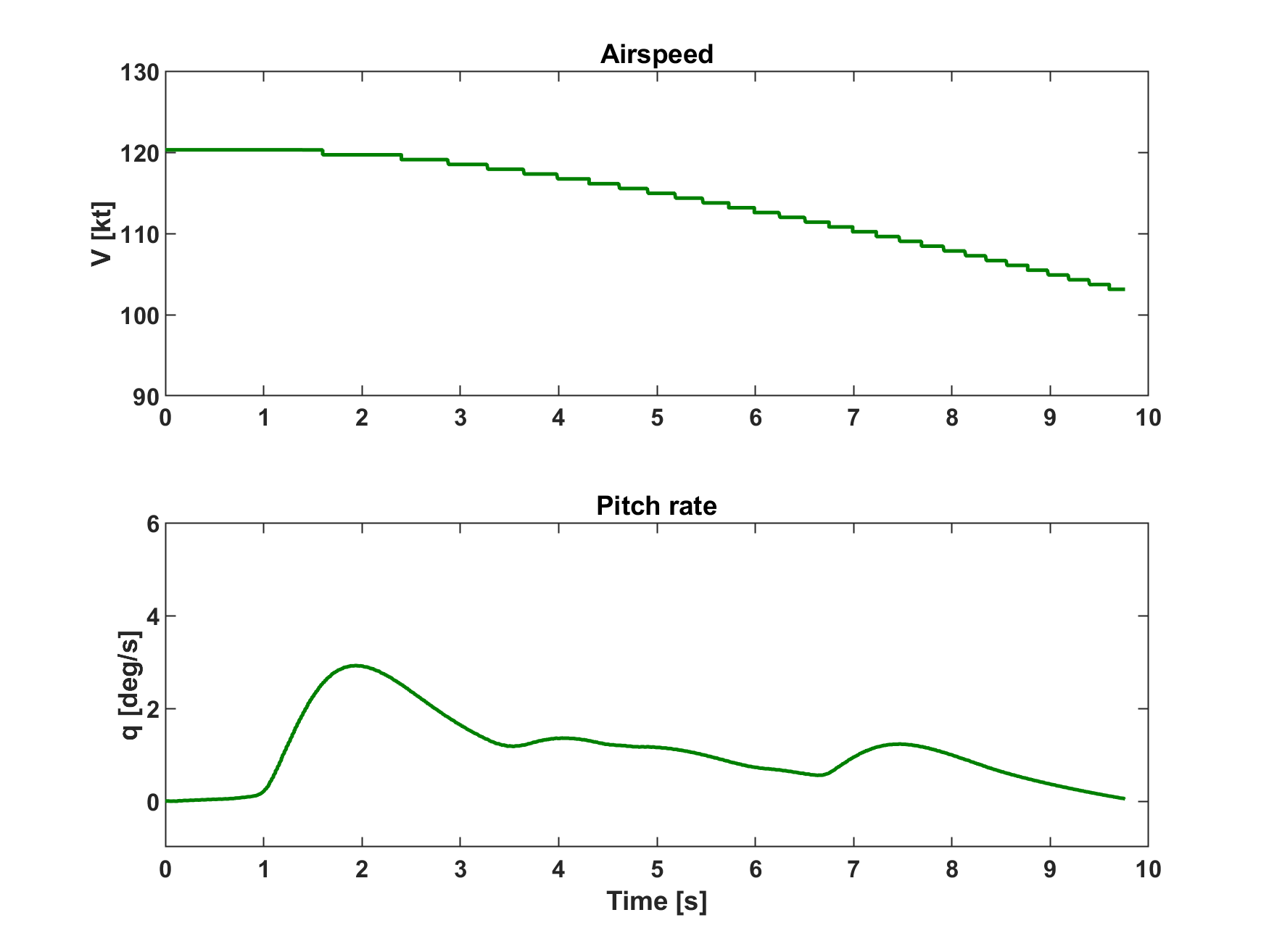}
    \caption{Cue on}
    \label{fig:right_b}
\end{subfigure}

\caption{Second experiment: vehicle response.}
\label{fig1:flight_test4}
\end{figure}

In this second experiment, the pilot is tasked with executing a longitudinal pulse input of 8 seconds duration and 20\% magnitude, both with the cue on and off.~The purpose of this experiment is twofold: to further assess the effectiveness of the cue in limiting the 1/rev load and, more importantly, to evaluate the pilot's ability to track the cue. Figures~\ref{fig1:flight_test3} and~\ref{fig1:flight_test4} showcase the results obtained from the experiment, displaying the 1/rev pitch link load of the nonlinear aircraft model, the pilot stick position, the computed control limit, and the vehicle pitch rate and airspeed.

In the cue-off scenario, the pilot exceeds the control limit significantly, leading to a substantial load exceedance. The aggressive pulse input induces a steep linear increase and decrease in pitch rate and airspeed, respectively, within the 1 to 5 seconds timeframe. Consequently, there is a sharp rise in the 1/rev load, resulting in a drastic decrease in the control limit from $10\%$ to $-30\%$.

In the cue-on scenario, the pilot begins tracking the cue at 1.7 seconds, recognizing the increasing aggressiveness of the input. The pilot adjusts to the cue within 0.3 seconds, after which almost perfect tracking is achieved. Although this adjustment time is efficient, it is justified by the inherent time delay in a visual cueing system. The pilot experiences a delay between perceiving the visual cue and initiating the corresponding sensory motor command.

It is noticeable that when the pilot utilizes the visual cue, the 1/rev load stays close to the load limit.~This reduction in load results in decreased vehicle maneuverability, as evidenced by the aircraft responses in the cue-on case. Despite achieving a significant reduction in load, it is crucial to highlight the load exceedance occurring in the cue-on case. This is primarily attributed to two factors.~First, a load estimation error arises from employing a simplified on-board model, and an additional approximation error stems from the linear representation of the 1/rev load of the nonlinear model.~Second, although the pilot accurately tracks the cue, there is occasional slight overshooting of the control limit. A potential avenue for research to improve the performance of the LLC scheme could involve adopting a control scheduling methodology instead of a fixed-point controller, where the LTI model would be scheduled with vehicle states, and the linear approximation of the 1/rev load would be updated during flight.

From the pilot's point of view, ``the visual cue was not intrusive'' as they could easily ignore it when needed to focus more on the mission.~The pilot also mentioned that ``the visual cue was easy to follow for slow to moderate control actuations. However, for more aggressive control actuations, it became harder to follow.''

The piloted simulations conducted thus far have primarily examined the use of the CM cue to reduce pilot control input aggressiveness. To understand the trade-off between load limiting and maneuver performance, it is essential for future piloted studies to incorporate ADS-33 maneuvers, such as the Pullup/Pushover maneuver. These maneuvers will also help assess the impact of the CM cue on the vehicle's handling quality. Given recent findings~\cite{ref_shah} that highlight the diverse impact of various harmonic loads on fatigue damage, future piloted simulation studies should also focus on integrating the LLC scheme with a Damage Mitigation Control (DMC) metric~\cite{ref_shah}. This integration will enable real-time tracking and limiting of the most damage-threatening harmonic loads. 

\section{Conclusion}

This paper introduces a novel life-extending control approach, the Load Limiting Control (LLC) scheme, designed for critical helicopter components. Synthesized using a higher-order linear time-invariant model of helicopter coupled body-rotor-inflow dynamics, the LLC scheme effectively resolves crucial issues inherent in existing life-extending control schemes. These issues include the difficulty in distinguishing between aggressive and non-aggressive maneuvers and the complete oversight of localized damage caused by specific harmonic loads.

The innovative features of the LLC scheme can be categorized into three main aspects. First, it expands the utilization of the LTI models of helicopter coupled body-rotor-inflow dynamics. Second, drawing inspiration from recent findings highlighting the diverse impact of various harmonic loads on fatigue damage, the proposed scheme can be customized to restrict the most damage-threatening harmonic load. Although the paper does not specifically delve into real-time determination of this critical load, a decision-making tool like the DMC metric could achieve this task. Finally, the proposed scheme trades maneuver performance for component load limiting only during aggressive maneuvers that induce significant fatigue damage, and at the pilot's discretion (if the LLC scheme is coupled with a cueing system). This stands in contrast to existing life-extending control schemes, like the Load Alleviation Control (LAC) scheme, where the trade-off between maneuver performance and load alleviation is always present, irrespective of maneuver aggressiveness.

To evaluate the effectiveness of the LLC scheme in limiting the magnitude of the 1/rev pitch link load, a comprehensive series of batch and real-time piloted simulations were conducted. The findings demonstrate that the LLC scheme effectively limits the 1/rev harmonic load within a predetermined limit, albeit with a discernible trade-off in maneuver performance. This trade-off is notably influenced by the chosen load limit and harmonic load. Despite successfully limiting the 1/rev load in the majority of instances, occasional load exceedances were noted and attributed to the utilization of a simplified dynamical model in constructing the LLC scheme. During the real-time piloted simulations, a Control Margin (CM) cue guided the pilot visually to adjust the control stick position for minimizing the 1/rev load. As evidenced by pilot feedback and data collected during flight tests, the pilot, after familiarizing with the cue during a training session, skillfully tracked the cue within half a second of deciding to initiate the tracking.

\section*{Acknowledgments}

This research was partially funded by the U.S. Government under Agreement No. W911 W6-17-2-0002.  The U.S. Government is authorized to reproduce and distribute reprints for Government purposes notwithstanding any copyright notation thereon.  The views and conclusions contained in this document are those of the authors and should not be interpreted as representing the official policies, either expressed or implied, of the U.S. Army Combat Capabilities Development Command Aviation and Missile Center (CCDC AvMC) or the U.S. Government.

\bibliography{sample}

\end{document}